\documentclass[12pt]{article}
\usepackage{mathrsfs,verbatim}
\usepackage{appendix}
\usepackage{natbib}
\usepackage{extarrows}
\usepackage{graphicx,setspace,lscape,longtable,epstopdf,xr}
\usepackage{mathrsfs,amsmath,amsthm,amssymb,color}
\usepackage{epsfig,graphicx}
\usepackage{rotating}
\usepackage{float}
\usepackage{bm}
\usepackage{ragged2e}
\usepackage{todonotes}
\usepackage{hyperref}
\usepackage{threeparttable}
\usepackage{multirow}
\usepackage{subfig}
\usepackage{dsfont}
\hypersetup{
colorlinks=true,
linkcolor=blue,
citecolor=blue,
urlcolor=blue}
\usepackage{booktabs}
\usepackage{algorithm}
\usepackage{algpseudocode}
\usepackage{amsmath}
\usepackage{graphicx}
\usepackage{hyperref}
\bibpunct{(}{)}{;}{a}{,}{,}

\newtheorem{theorem}{Theorem}
\newtheorem{lemma}{Lemma}
\newtheorem{proposition}{Proposition}

\newtheorem{assumption}{Assumption}
\def\one{{\mathbf{1}}}
\def\bA{{\mathcal A}}
\def\ve{\varepsilon}

\def\tr{\mbox{tr}}

\def\n{\nonumber}

\def\beq{\begin{equation}}
\def\eeq{\end{equation}}
\def\beqr{\begin{eqnarray}}
\def\eeqr{\end{eqnarray}}
\def\beqrs{\begin{eqnarray*}}
\def\eeqrs{\end{eqnarray*}}
\def\bet{\begin{theorem}}
\def\eet{\end{theorem}}
\def\bel{\begin{lemma}}
\def\eel{\end{lemma}}
\def\bep{\begin{proposition}}
\def\eep{\end{proposition}}
\def\bg{\begin{figure}[tbph]\begin{center}}
\def\eg{\end{center}\end{figure}}

\def\bc{\begin{center}}
\def\ec{\end{center}}

\def\e{\mathbf{e}}

\newtheorem{remark}{Remark}
\newtheorem{corollary}{Corollary}

\def\wt{\widetilde}
\def\cle{\preccurlyeq}

\def\wh{\widehat}

\def\ol{\overline }

\def\E{\mathbb{E}}
\def\bB{\mathbf{B}}

\def\mD{\mathcal D}

\def\mR{\mathbb{R}}

\def\mL{\mathcal L}

\def\mF{\mathcal F}

\def\bR{\mathbf{R}}
\def\bY{\mathcal Y}

\def\mE{\mathcal E}

\def\bH{\mathbb H}

\def\bZ{\mathbf{Z}}
\def\bV{\mathbf{V}}
\def\bU{\mathbf{U}}

\def\var{\mbox{var}}

\def\cov{\mbox{cov}}

\def\diag{\mbox{diag}}

\def\vec{\mbox{vec}}

\def\bgamma{\boldsymbol{\gamma}}
\def\bbeta{\boldsymbol{\beta}}

\def\bLambda{\boldsymbol{\Lambda}}
\def\bSigma{\boldsymbol{\Sigma}}

\newcommand{\RNum}[1]{\uppercase\expandafter{\romannumeral #1\relax}}
\def\bX{\mathbf{X}}
\def\bZ{\mathbf{Z}}

\def\bY{\mathbf{Y}}
\def\bA{\mathbf{A}}

\def\bD{\mathbf{D}}
\def\bW{\mathbf{W}}

\def\bH{\mathbf{H}}
\def\b{\mathbf{b}}

\def\bg{\mbox{\boldmath $g$}}
\def\bI{\mathbf{I}}

\def\lag{\rm lag}

\def\zero{\mathbf{0}}
\def\defeq{\stackrel{\mathrm{def}}{=}}  

\renewcommand{\arraystretch}{1.3}

\numberwithin{equation}{section}

\def\be{\begin{eqnarray*}}
\def\ese{\end{eqnarray*}}
\def\be{\begin{eqnarray}}
\def\ee{\end{eqnarray}}
\def\bsq{\begin{equation*}}
\def\esq{\end{equation*}}
\def\bq{\begin{equation}}
\def\eq{\end{equation}}

\def\var{\hbox{var}}
\def\cov{\hbox{cov}}

\def\wh{\widehat}

\def\wt{\widetilde}

\def\mL{\mathcal L}
\def\mR{\mathbb{R}}
\def\n{\nonumber}

\def\cov{\mbox{cov}}

\def\vec{\mbox{vec}}

\def\diag{\mbox{diag}}
\def\tr{\mbox{tr}}

\def\trans{^{\top}}

\def\bGamma{\boldsymbol\Gamma}

\def\bg{\boldsymbol\gamma}

\def\0{{\bf 0}}
\def\1{{\bf 1}}
\def\A{{\bf A}}

\def\e{{\bf e}}

\def\a{{\bf a}}
\def\B{{\bf B}}

\def\r{{\bf r}}

\def\b{{\bf b}}
\def\I{{\bf I}}

\def\M{{\bf M}}
\def\bM{{\bf M}}

\def\bP{{\bf P}}
\def\bP{{\bf P}}

\def\bV{{\bf V}}

\def\W{{\bf W}}

\def\X{{\bf X}}

\def\I{{\bf I}}

\def\Y{{\bf Y}}

\def\Z{{\bf Z}}

\def\bSig{{\bf \Sigma}}

\def\diag{\hbox{diag}}

\def\bq{\begin{equation}}
\def\eq{\end{equation}}

\def\wh{\widehat}
\def\wt{\widetilde}

\def\diag{\hbox{diag}}
\def\log{\hbox{log}}

\def\squarebox#1{\hbox to #1{\hfill\vbox to #1{\vfill}}}
\def\btheta{{\boldsymbol \theta}}
\def\bfeta{{\boldsymbol \eta}}
\def\balpha{{\boldsymbol \alpha}}

\def\blambda{{\boldsymbol \lambda}}

\def\bpi{{\boldsymbol \pi}}

\def\vec{\mathrm{vec}}

\def\bE{\mathbf{E}}

\def\mF{\mathcal{F}}

\def\var{\hbox{var}}
\def\cov{\hbox{cov}}

\def\bse{\begin{eqnarray*} }
\def\ese{\end{eqnarray*}}
\def\be{\begin{eqnarray}}
\def\ee{\end{eqnarray}}
\def\bsq{\begin{equation*}}
\def\esq{\end{equation*}}
\def\bq{\begin{equation}}
\def\eq{\end{equation}}

\def\br{{\bf r}}

\def\wh{\widehat}
\def\wt{\widetilde}
\def\diag{\hbox{diag}}

\def\diag{\hbox{diag}}

\def\boxit#1{\vbox{\hrule\hbox{\vrule\kern6pt\vbox{\kern6pt#1\kern6pt}\kern6pt\vrule}\hrule}}

\begin{document}
\begin{center}

{\bf\Large  Network Autoregression for Incomplete Matrix-Valued Time Series}\\

\bigskip
		Xuening Zhu$^1$, Feifei Wang$^{2*}$, Zeng Li$^3$ and Yanyuan Ma$^4$

		{\it $^1$Fudan University,  China;
			$^2$Renmin University of China, China;\\
            $^3$Southern University of Science and Technology, China;\\
            $^4$The Pennsylvania State University, USA
}

\end{center}

\begin{footnotetext}[1]{Feifei Wang is the corresponding author.}
\end{footnotetext}

\begin{singlespace}
		\begin{abstract}
 		
We study the dynamics of matrix-valued time series with observed
network structures by proposing a matrix network autoregression model
with row and column networks of the subjects.
We incorporate covariate information and a low rank intercept
matrix. We allow incomplete observations in the matrices and the
missing mechanism can be covariate dependent.
To estimate the model, a two-step estimation procedure is proposed.
The first step aims to estimate the network autoregression
coefficients, and the second step aims to estimate the regression
parameters, which are matrices themselves.
Theoretically, we first separately establish the asymptotic properties of the
autoregression coefficients and the error bounds of the regression
parameters.
Subsequently, a bias reduction procedure is proposed to reduce the
asymptotic bias and the theoretical property of the debiased estimator
is studied.
Lastly, we illustrate the usefulness of the proposed method through a
number of
numerical studies and an analysis of a Yelp data set.\\

\noindent {\bf KEY WORDS: }  Bias reduction; Incomplete matrix observations;
Matrix-valued time series;
Network autoregression.

		\end{abstract}
	\end{singlespace}

\newpage
\section{Introduction}

Due to the improved data collection capability, matrix-valued time
series data have become increasingly popular in various fields, such as
economics, finance, environmental sciences and many others
\citep{leng2012sparse,zhou2014regularized,zhou2014gemini,wang2019factor,chen2021factor}.
Examples include: the import-export volumes among a group of countries
within a time period, the environmental indicators (e.g., PM$_{2.5}$,
SO$_2$, CO, NO$_{2}$)
recorded by air pollution monitoring stations during a year, and the economic indicators (e.g., GDP, CPI, basic interests rate, unemployment rate) published by different countries every year.
Given the wide applicability of matrix-valued time series data, it is of great importance to understand the inner dynamic patterns of this type of data for better practical applications.


To model the dynamics of the matrix-valued data, a straightforward
method is to
stack it into a long vector and use standard time series analysis
tools for vector form data \citep{lutkepohl2005new}.
However, this approach ignores the inner relationships between the rows
and columns of the matrix and lacks interpretability
\citep{chen2021factor}.
Moreover, it increases the number of model parameters to be estimated and
therefore results in inferior model performances.
To enhance model interpretability and reduce model parameters,
\cite{chen2021autoregressive}
propose a matrix autoregression model with a bilinear form.
\cite{wang2021high} extend the matrix autoregression to tensor
autoregression model and utilize a low-rank Tucker decomposition for
dimension reduction.
To model the cross-sectional time dependence of the matrix-valued time
series,
\cite{wang2019factor}, \cite{chang2021modelling}, \cite{chen2021statistical},
\cite{yu2021projected}, and \cite{kong2022matrix}
 consider dynamic factor and low rank
structures for dimension reduction.
Despite of their usefulness, the matrix-valued time series
  models still have some issues.
First, although the matrix autoregression model of
\cite{chen2021autoregressive} can significantly
reduce the model parameters when compared with the vector model approach, it still needs to estimate $O(N^2)$ parameters for an $N\times N$ matrix-valued time series.
This results in suboptimal estimation results, especially when $N$ is extremely large.
Second, although factor models can characterize the unobserved
dependence structure, it does not take  the observed dependence
structure (such as
the network structure) into consideration.
Third, current model estimation approaches for matrix data typically
assume that all entries in the matrix are
completely observed. These methods cannot be directly applied for incomplete
matrix-valued data, which are also frequently encountered in practice
\citep{cai2016structured,bi2017group,mao2019matrix}.

Motivated by the above issues, we propose a network autoregression
model for
incomplete matrix-valued time series data.
First, we take advantage of the observed network structure among the
subjects into the modelling process.
For example, suppose we collect matrix-valued data
from individuals who visit shops located in several spatial regions
and aim to characterize the users' dynamic visiting behaviors for user
profiling analysis.
In this case, we can collect the social network among the individuals
as well as
the spatial network among the regions.
The two types of networks are referred to as row and column
networks, respectively.
These network information can help us to predict the individual
visiting behavior
by considering his/her connected friends or neighboring regions.
Therefore we explicitly embed the observed network structure information into our modelling process.
Meanwhile, this allows us to further reduce the number of parameters
using the approach of \cite{zhu2017network}.

Second, we consider model estimation for incomplete
matrix-valued time series data.
Matrix completion problem has been considered recently in the
literature under
multiple missing mechanisms
\citep{koltchinskii2011nuclear,rohde2011estimation,cai2016structured,mao2019matrix}.
However, these works typically focus on the static data setting. In this work, we develop a network autoregression model for matrix-valued time series to investigate its dynamic patterns.
Motivated by the matrix completion literature, we assume the
intercept (matrix) in the matrix autoregression to be low rank.
In addition, we utilize
  the time-invariant covariate information, which also enters into the
  model in a linear form with a possibly diverging dimension.
This allows us to conduct dynamic matrix completion as we illustrate
in our empirical study.

Our model estimation is conducted in two steps.
In the first step, we use a profile objective function to estimate
the autoregressive coefficients.
Specifically, we use a logistic regression to model the missing mechanisms
and take the inverse probability in the profile object function.
In the second step,
we estimate the regression coefficients of the
covariates along with the low rank intercept matrix with a separate projection
procedure and singular value decomposition (SVD).
Using a designed algorithm, our model estimation is computationally tractable and efficient. It is unconventional to split the estimation for the two groups of parameters into two steps.
Theoretically, we first establish the asymptotic properties for the
network autoregressive coefficient estimator.
Furthermore, a bias reduction procedure is proposed to reduce the
estimation bias
of the estimator.
Finally, we investigate the error bounds of the debiased estimator.
Different from the static data setting, the dynamic and network
dependence information is  further considered in the theoretical
analysis. Extensive numerical studies and a real data example from
Yelp ({\it www.yelp.com}) are then used to illustrate the proposed
methodology.

The article is organized as follows.
Section \ref{sec:model_notations} introduces the model and notations.
Section \ref{sec:model_est} presents model estimation procedure.
Section \ref{sec:theory} establishes theoretical properties of the
proposed model.
Section \ref{sec:numerical} conducts a variety of numerical studies to
evaluate the finite sample performance of the method. Section \ref{sec:real_data} presents an analysis on a Yelp
  data set. We conclude in Section \ref{sec:conclude}.
All technical proofs are relegated to the supplementary file.

\section{Model and Notations}\label{sec:model_notations}

Let $\bY_t = (Y_{ijt})\in \mR^{N_1\times N_2}$
be a dynamic high-dimensional matrix of interest
containing joint information of two aspects.
For instance, $Y_{ijt}$ is the response (e.g., the number of visits) of the
$i$th user in the
$j$th  spatial region at the $t$th time point.
In practice, we assume only a portion of responses are observed.
To denote the missing mechanisms,
we use a binary indicator $R_{ijt}\in\{0,1\}$,
where $R_{ijt} = 1$ implies that
the $(i,j)$th response is observed at the $t$th time point, and $R_{ijt} = 0$ otherwise.
We assume that $R_{ijt}$ follows a Bernoulli distribution with
parameter $P(R_{ijt}=1) = p_i$.
Here we assume $p_i$ depends on some covariate information of
   the $i$th user, which is denoted by
    $\X_{i}\in\mR^p$.
Specifically,
we assume a logistic regression model for  $R_{ijt}$, i.e.,
\begin{align}
P(R_{ijt} = 1) =  p_{i} = \frac{\exp\big(\bX_{i}^\top \balpha \big)}{1+\exp\big(\bX_{i}^\top \balpha \big)},\label{logistic_R}
\end{align}
for $t=1, \dots, T$, where $\bm{\alpha}$ denotes the corresponding
coefficients for $\bX_{i}$. For convenience, let $\bX\equiv(\X_{1},
\dots, \X_{N_1})\trans\in \mR^{N_1\times p}$.

Besides the covariate information of the users, we can also observe
the network relationships
among the users. We record the network information by $\bA_1 = (a_{1ij})\in \mR^{N_1\times N_1}$,
where $a_{1ij} = 1$ indicates that the $i$th user  is connected to the
$j$th user, otherwise $a_{1ij} = 0$.
Similarly,
the spatial relationships of regions can also be constructed by $\bA_2 = (a_{2ij})\in \mR^{N_2\times N_2}$, where $a_{2ij} = 1$ denotes that
 the $i$th region is a spatial neighbor to the $j$th region, and $a_{2ij} = 0$ otherwise.
Following the convention, we set $a_{1ii} = 0$ for $1\le i\le N_1$ and
$a_{2jj} = 0$ for $1\le j\le N_2$.
In general, we name $\bA_1$ as the \emph{row network} of $N_1$
row subjects, and refer to $\bA_2$ as the \emph{column network} of $N_2$
column  locations.
Of course, the row and column network nodes are not necessarily
users and locations in real life. They can be any subjects
whenever the row and column networks can be appropriately constructed.
Define $\bW_1 \equiv (d_{1i}^{-1}a_{1ij})\in \mR^{N_1\times N_1}$ as
the row-normalized adjacency matrix for $\bA_1$
with $d_{1i} = \sum_j a_{1ij}$.
Similarly define $\bW_2 = (d_{2j}^{-1}a_{2ij})\in \mR^{N_2\times
  N_2}$ with $d_{2j} = \sum_i a_{2ij}$. In other words, $\bW_2$ is a
column-normalized adjacency matrix for $\A_2$.
In our analysis, we allow $N_1, N_2$ to diverge, and assume
  $N_1\asymp N_2$ for technical convenience.
Define $m = N_1+N_2$.

In this article we investigate
how to  conduct matrix-valued network autoregression with incomplete
matrix entries.
We consider the following matrix network autoregression (MNAR) model,
\beq
\bY_t = \bLambda \bW_1\bY_{t-1} + \bY_{t-1}\bW_2\bGamma + \bX\bbeta  + \bB + \mE_t,\label{model}
\eeq
where $\bLambda = \diag\{\lambda_{i}:1\le i\le N_1\}\in \mR^{N_1\times N_1}$,
$\bGamma = \diag\{\gamma_{i}:1\le i\le N_2\}\in \mR^{N_2\times N_2}$,
$\bbeta\in \mR^{p\times N_2}$, $\bB=(b_{ij})\in \mR^{N_1\times N_2}$
are unknown parameter matrices,
and $\mE_t = (\ve_{ijt})\in \mR^{N_1\times N_2}$ is the noise matrix.
We assume $\var(\ve_{ijt})=\sigma^2$, all $\{\ve_{ijt}\}$s are
  independent of each other and also independent of $\Y_{t-1}$.
Following the literature \citep{rohde2011estimation,mao2019matrix,wang2021high}, we impose a low-rank structure on the intercept matrix
$\bB$. Specifically, we assume
$\bB = \bU\bV^\top$, where $\bU\in \mR^{N_1\times r_B}$, $\bV\in \mR^{N_2\times r_B}$, and $r_B$ is the rank of $\bB$ with
$r_B\ll \min(N_1,N_2)$.
To guarantee identifiability, we assume the column spaces of $\bB$
and $\bX$ are orthogonal to each other, i.e., $\bX\trans\bB=\0$.

\begin{remark}
  The first two terms in the right hand side of (\ref{model})
  characterize the local network effects from connected users and
  locations.
  It is a generalization of the network/spatial autoregression model
  \citep{zhu2017network} to the high-dimensional matrix data.
  To better understand the model, we can consider the special case
  where $\bGamma = \zero$.
Then we obtain
\begin{align*}
\bY_t = \sum_{k = 0}^\infty \bLambda^k\bW_1^k \Big(\bX\bbeta + \bB + \mE_{t-k}\Big).
\end{align*}
Hence the proposed model incorporates a user's own historical
information as well as that of its connected neighbors.
  If $\bLambda = \zero$ and $\bGamma = \zero$,
  then $\bY_t$ is fully characterized by the structure of a
    covariate effect plus a low rank matrix. In this case, the problem is reduced to the setting of matrix completion problem with
  exogenous covariates \citep{mao2019matrix}.
\end{remark}

\begin{remark}
Recently, \cite{chen2021autoregressive} considered a bilinear
autoregression model for matrix-valued time series. A direct
extension to the network framework is
  \begin{align}
    \bY_t = \bLambda \bW_1\bY_{t-1}\bW_2\bGamma + \bX\bbeta  + \bB + \mE_t.\label{eq:mnar_bilinear}
  \end{align}
This is different from our model (\ref{model}) in
that the network effects $\bLambda$ and $\bGamma$ in (\ref{eq:mnar_bilinear})
act in a multiplicative fashion.
  Similar multiplicative effect has been considered by
  \cite{wu2021inward} to characterize the inward and outward
  influences of the network nodes in vector data.
In contrast, we consider the network effects in an additive form. This
modeling difference leads to completely different estimation
procedure and theoretical development.
\end{remark}

{\sc General Notations.}
For a symmetric or Hermitian matrix $\bA$,
we use $\lambda_{\max}(\bA)$ and $\lambda_{\min}(\bA)$ to denote
its maximum and minimum eigenvalues, respectively.
For an arbitrary matrix $\M = (M_{ij})\in \mR^{m\times n}$, define
$\sigma_1(\M)$ as the largest singular value of $\M$,
and $\rho(\bM)$ as the spectral radius of $\bM$.
Define $\|\bM\|_F = \tr(\bM^\top\bM)^{1/2}$ as the Frobenius norm of the matrix $\bM$.
Let $\M_{i\cdot}\in \mR^{1\times n}$
and $\M_{\cdot j}\in \mR^{m}$
denote the $i$th row vector or the $j$th column vector of $\M$, respectively.
Let $\vec(\bM)\in \mR^{mn}$ be the vectorization of $\bM$ by stacking
its column vectors into a long vector.
Let $|\M|_e = (|M_{ij}|)\in \mR^{m\times n}$, where
we take absolute value of each element in $\M$.
In addition, define $\M_{1}\cle \M_2$ if $M_{1,ij}\le M_{2,ij}$
for $1\le i\le m, 1\le j\le n$.
Let $\{a_n\}$ and $\{b_n\}$ be two sequences related to $n$.
 Define $a_n\lesssim b_n$ as $a_n\le cb_n$ for some constant $c$ as $n\to \infty$.
Define $a_n\ll b_n$ when $a_n/b_n\to 0$  as $n\to \infty$.
Let $\e_i\in \mR^n$ be the vector with the $i$th element being 1 and the
others being 0.
Define $\one_n\in \mR^n$ as an $n$-dimensional vector  whose elements are all one.
Let $\bI_n\in \mR^{n\times n}$ be the $n$-dimensional identity matrix.
In addition, let $[N] = \{1,2,\cdots, N\}$ for any integer $N$.

{\sc Notation of Norms.}
For a vector $\a$, we use $\|\a\|$ to denote its $l_2$ norm. For any matrix $\M$, define $\|\M\|_{\max} = \max_{i,j}|M_{ij}|$ and
$\|\bM\| = \sigma_1(\bM)$. Further denote $\|\bM\|_* = \sum_k \sigma_k(\bM)$ as the nuclear norm of the
matrix $\bM$.

\section{Model Estimation}\label{sec:model_est}

\subsection{Two-Step Estimation Procedure}

To estimate the unknown parameters in MNAR, we first assume the matrix entries of the response
are fully observed, and consider a least squares type objective
function. Then we derive a profile objective function for incomplete
matrix data and develop a two-step estimation procedure. Let
$\btheta = (\blambda^\top, \bgamma^\top, \vec(\bbeta)^\top, \vec(\bB)^\top)^\top$
denote all parameters to be
estimated,
where $\blambda = \diag(\bLambda)$ and $\bgamma = \diag(\bGamma)$.
If we could observe the whole matrix $\bY_t$, then we could
minimize the following least squares objective function
\begin{align}
Q^F(\btheta) = \sum_{t = 1}^T\Big\|
 \bY_t - \bLambda \bW_1  \bY_{t-1} -
 \bY_{t-1}\bW_2\bGamma - \bX\bbeta  - \bB
\Big\|_F^2\label{Q_F}
\end{align}
to obtain the parameter $\btheta$, where $F$ stands for \emph{full data}.
Define
\begin{align*}
\Delta_{ijt}^F(\btheta) =
Y_{ijt} -
\lambda_i\sum_{k=1}^{N_1} {W_{1ik}Y_{kj(t-1)} }
- \sum_{k=1}^{N_2} {Y_{ik(t-1)}W_{2kj}\gamma_j}
- \xi_{ij},
\end{align*}
where $\xi_{ij} = \bX_{i}^\top \bbeta_{\cdot j}  +b_{ij}$.
Then we have $Q^F(\btheta) = \sum_{i,j,t} \Delta_{ijt}^{F}(\btheta)^2$.

Although the objective function (\ref{Q_F}) is straightforward, it
cannot be directly applied due to two main issues.
The first issue is the missing values of the matrix entries. In fact,
not all entries of $\bY_t$ are observable.
To handle the missingness in $\bY_t$, we define $\bZ_t =
(Z_{ijt})\in\mR^{N_1\times N_2}$,
where $Z_{ijt} = R_{ijt}Y_{ijt}/ p_i$
 is the inverse probability weighted response.
We then devise the estimation method based on the $\bZ_t$ matrix
in the presence of incomplete matrix entries.
The second issue is the complex form of the objective function,
which contains different parameters, thus requiring different computational
treatments and having different theoretical properties.
For example, to take into account the model properties, we need
to impose several penalties on a subset of the parameters, which makes
it hard to optimize with respect to all parameters simultaneously, and can lead
to low computational efficiency.
Moreover, from a theoretical aspect,
the statistical convergence rates of the parameters are inherently
different, while analyzing a simultaneous optimization
procedure is hard  and  may result in artificially lowered
convergence rate of some parameters.

To address the two concerns, we devise a two-step estimation procedure. In the first step, we estimate
the network effects ($\bLambda$ and $\bGamma$) by minimizing an objective
function only involving the parameters $\bLambda$ and $\bGamma$. In the second step, the regression matrix parameters ($\bbeta$ and $\bB$) are estimated by minimizing a different objective function with the estimates of $\bLambda$ and $\bGamma$ plugged in.
The two-step procedure leads to
a computationally efficient algorithm for model estimation. Additionally,
separating the estimation of parameters in such a two-step procedure can facilitate natural and convenient
theoretical investigation, and ensure both sets of  parameters
achieve their proper convergence rates. In the following, we present the two-step estimation procedure in detail.
Define
\begin{align}
\Delta_{ijt}(\btheta) &=
Z_{ijt} -\lambda_i\sum_{k  =1}^{N_1} W_{1ik} Z_{kj(t-1)}
- \sum_{k  =1}^{N_2} Z_{ik(t-1)}W_{2kj}\gamma_j
- \xi_{ij}\nonumber\\
&=Z_{ijt} -
\lambda_i\W_{1i\cdot}\Z_{\cdot j(t-1)}
- \Z_{i\cdot(t-1)}\W_{2\cdot j}\gamma_j
- \xi_{ij}.\label{Delta_ijt}
\end{align}
We first focus on the estimation of the network effects, i.e.,
$\blambda = (\lambda_1,\cdots, \lambda_{N_1})^\top$ and
$\bgamma = (\gamma_1,\cdots, \gamma_{N_2})^\top$.
To this end, we profile out $\xi_{ij}$ by forming
\bse
\wt\Delta_{ijt}(\btheta) &=& \Delta_{ijt}(\btheta) - \ol
\Delta_{ij}(\btheta)\\
&=&Z_{ijt} -\ol Z_{ij}-
\lambda_i\W_{1i\cdot}(\Z_{\cdot j(t-1)}-\ol\Z_{\cdot j,\lag})
- (\Z_{i\cdot(t-1)}-\ol\Z_{i\cdot,\lag})
\W_{2\cdot j}\gamma_j\\
&=&\wt Z_{ijt} -
\lambda_i\W_{1i\cdot}\wt \Z_{\cdot j(t-1),\lag}
- \wt\Z_{i\cdot(t-1),\lag}\W_{2\cdot j}\gamma_j,
\ese
where $\ol \Delta_{ij}(\btheta) = T^{-1}\sum_{t=1}^{T}\Delta_{ijt}(\btheta)$,
$\ol \bZ = T^{-1}\sum_{t=1}^{T}\bZ_t = (\ol Z_{ij})$ and $\ol \bZ_{\lag} = (T-1)^{-1}\sum_{t=2}^{T}\bZ_{t-1} =(\ol Z_{ij,\lag})$.
Let $\wt\bZ_t = \bZ_t - \ol \bZ = (\wt Z_{ijt})$ and $\wt \bZ_{t-1,\lag} = \bZ_{t-1} - \ol\bZ_{\lag} = (\wt Z_{ij(t-1), \lag})$.
It is easy to verify that $E\{\wt\Delta_{ijt}(\btheta)|\mD\} =
\wt\Delta_{ijt}^F(\btheta)$,
where $\mD = \{(\bX_t, \bY_t):1\le t\le T\}$ and $\wt \Delta_{ijt}^F(\btheta) = \Delta_{ijt}^F(\btheta) - \ol \Delta_{ij}^F(\btheta)$.
However, we point out that  $E\{\wt\Delta_{ijt}(\btheta)^2|\mD\} \ne
\wt\Delta_{ijt}^{F}(\btheta)^2$,
because  $E(Z_{ijt}^2|\mD)= p_i^{-1}Y_{ijt}^2\ne Y_{ijt}^2$.
Consequently we cannot directly use $\wt\Delta_{ijt}(\btheta)^2$
to replace
$\wt\Delta_{ijt}^F(\btheta)^2$ in the objective function.
Noting $\sum_t\wt\Delta_{ijt}(\btheta)^2
 = \sum_t\Delta_{ijt}(\btheta)^2 - T\ol \Delta_{ij}(\btheta)^2$,
 we modify $\sum_t\wt\Delta_{ijt}^2(\btheta)$ to
\begin{align*}
\wt\Delta_{ij}^{c2}(\btheta) &= \sum_t\wt\Delta_{ijt}^2(\btheta)
+\lambda_i^2(1-T^{-1})\sum_t\sum_{k = 1}^{N_1}W_{1ik}^2Z_{kj(t-1)}\Big(Y_{kj(t-1)} - Z_{kj(t-1)}\Big) \\
&+
\gamma_j^2(1-T^{-1})\sum_t\sum_{k = 1}^{N_2}W_{2kj}^2Z_{ik(t-1)}\Big(Y_{ik(t-1)} -
                               Z_{ik(t-1)}\Big)
+(1-T^{-1})\sum_tZ_{ijt}(Y_{ijt} - Z_{ijt}),
\end{align*}
where the superscript $c$ stands for \emph{correction}.
We can verify that $E\{\wt\Delta_{ij}^{c2}(\btheta)\}
 = \sum_tE\{\wt\Delta_{ijt}(\btheta)^2\}$.
Based on the above analysis, let $\wt Q(\btheta) = \sum_{i,j}
  \wt \Delta_{ij}^{c2}(\btheta)$
be the profile objective function.
We also note that $\blambda$ and $\bgamma$ are of high
dimensionality due
to the increasing network sizes $N_1$ and $N_2$.
Therefore it is natural to take penalization method into consideration.
Specifically, we consider to minimize
the following profile objective function with ridge penalization
\beq
\wt Q_p(\btheta)=\wt Q(\btheta) + \nu_1\|\bLambda\|_F^2 +
\nu_2\|\bGamma\|_F^2,\label{Q_obj}
\eeq
where $\nu_1$ and $\nu_2$ are two tuning parameters. Let $(\wt \bLambda, \wt \bGamma) = \arg\min_{\wt \bLambda, \wt \bGamma} \wt Q_p(\btheta)$ be the penalized profile estimator.

Next, with the penalized profile estimator $(\wt \bLambda, \wt \bGamma)$,  we proceed to estimate $\bbeta$ and $\bB$ in the second step. Define
\begin{align}
\wt\bE_t = \bZ_t - \wt\bLambda\bW_1\bZ_{t-1} - \bZ_{t-1}\bW_2\wt\bGamma\label{Et}
\end{align}
as the residual. Then we treat $\wt\bE_t$ as the response to estimate
$\{\bbeta, \bB\}$.
Due to the high-dimensionality of the parameters, we follow
\cite{mao2019matrix} to consider the
following penalized objective function,
\begin{align}
Q_p^{(2)}(\btheta) =  \big\|T^{-1}\sum_{t = 1}^T\wt\bE_t - \bX\bbeta - \bB\big\|_F^2 +
\nu_3\|\bbeta\|_F^2+
\nu_4\Big(\alpha\|\bB\|_*+(1-\alpha) \|\bB\|_F^2\Big),\label{Q_p2}
\end{align}
where $\|\bB\|_*$ denotes the nuclear norm of $\bB$, $\nu_3$, $\nu_4$ and $\alpha$ are tuning parameters.
By using the penalization term $\nu_3\|\bbeta\|_F^2$ and $ \nu_4(1-\alpha) \|\bB\|_F^2$, we aim to achieve
$L_2$-shrinkage of $\bbeta$ and $\bB$, which is computationally efficient and
helpful in dealing with high-dimensional problems.
In addition, the nuclear penalization $\nu_4\alpha\|\bB\|_*$ is used
to penalize the singular values of $\bB$ to encourage a low rank
structure.
The tuning parameter $\alpha \in [0,1]$ is set to strike a balance
between the nuclear penalization and
$L_2$-shrinkage on $\B$.
All tuning parameters are selected with 5-fold cross-validation method in our numerical studies.

\subsection{An Iterative Optimization Algorithm}

We discuss the implementation and optimization algorithm for the two-step estimation method in this section.
Note that the penalized profile objective function (\ref{Q_obj})
has a least squares form, thus the estimator can be obtained
analytically.
However, one can note that the derivation of its analytical form involves
an inverse of a high-dimensional matrix of dimension $m\times m$, where
$m = N_1+N_2$.
Therefore, it poses huge computational challenges especially for large
scale networks.

To solve this issue, we develop an iterative algorithm for model estimation. It is notable that, the parameter $\blambda$ or $\bgamma$ can be easily obtained once we fix the other as known.
Take $\bgamma$ as fixed for example.
Under this case, we can estimate $\blambda$ by minimizing
$\wt Q(\btheta) +
    \nu_1\|\bLambda\|_F^2$.
Define $\Delta_{1it}^{\gamma} = (\wt\bZ_{i\cdot t}-\wt\bZ_{i\cdot (t-1),\lag}\bW_2
    \bGamma)\trans \in \mR^{N_2}$,
  and then we obtain
    \begin{align}
    \wt \lambda_{i} = \Big(\sum_t \bW_{1i\cdot}\wt\bZ_{t-1,\lag}
    \wt\bZ_{t-1,\lag}^\top \bW_{1i\cdot}^\top+ \kappa_{i}+\nu_1 \Big)^{-1}\Big(\sum_t \bW_{1i\cdot}\wt\bZ_{t-1,\lag} \Delta_{1it}^{\gamma}\Big),\label{lamb_fixed_gamma}
    \end{align}
    where $\kappa_i = (1-T^{-1}){\sum_t}\sum_{j=1}^{N_2}\sum_{k =
      1}^{N_1}W_{1ik}^2Z_{kj(t-1)}\big(Y_{kj(t-1)} -
    Z_{kj(t-1)}\big)$.
Particularly, we note that (\ref{lamb_fixed_gamma}) can be obtained for each
$\lambda_i$ with $1 \leq i \leq N_1$ separately, when given $\bgamma$.
Consequently, it does not involve the inverse of a high-dimensional matrix, which makes it computationally friendly.
Similarly, when $\blambda$ is given, we can derive the analytical forms for $\wt \gamma_j$ with $1 \leq j \leq N_2$.
This leads to the iterative algorithm as follows.

\begin{itemize}
  \item [Step 1.] Obtain initial estimates $\bLambda^{(0)}$ and
  $\bGamma^{(0)}$.
  \item [Step 2.] Let $M$ define the number of iterations required for convergence. For $m = 1,\cdots M$, denote the $m$th estimator as
  $\bLambda^{(m)}$ and
  $\bGamma^{(m)}$.
  Then repeat the following Steps 2.1--2.2 until convergence.
  \begin{itemize}
    \item [Step 2.1.] Given $\bGamma^{(m)}$, optimize $\bLambda$ by minimizing $\wt Q(\btheta) +
    \nu_1\|\bLambda\|_F^2$.
    Define $\Delta_{1it}^{(m)} = (\wt\bZ_{i\cdot t}-\wt\bZ_{i\cdot (t-1),\lag}\bW_2
    \bGamma^{(m)} )\trans \in \mR^{N_2}$,
  and then we obtain
    \begin{align*}
    \lambda_{i}^{(m+1)} = \Big(\sum_t \bW_{1i\cdot}\wt\bZ_{t-1,\lag}
    \wt\bZ_{t-1,\lag}^\top \bW_{1i\cdot}^\top+ \kappa_{i}+\nu_1 \Big)^{-1}\Big(\sum_t \bW_{1i\cdot}\wt\bZ_{t-1,\lag} \Delta_{1it}^{(m)}\Big),
    \end{align*}
    where $\kappa_i = (1-T^{-1}){\sum_t}\sum_{j=1}^{N_2}\sum_{k =
      1}^{N_1}W_{1ik}^2Z_{kj(t-1)}\big(Y_{kj(t-1)} -
    Z_{kj(t-1)}\big)$.
    \item [Step 2.2.]
    Given $\bLambda^{(m+1)}$, optimize $\bGamma$ by minimizing $\wt Q(\btheta) + \nu_2\|\bGamma\|_F^2$.
    Define $\Delta_{2jt}^{(m)} = \wt\bZ_{\cdot j t}-
    \bLambda^{(m+1)}\bW_1\wt\bZ_{\cdot j(t-1),\lag} \in \mR^{N_1}$,
 and then we obtain
    \begin{align*}
    \gamma_{j}^{(m+1)} = \Big(\sum_t \bW_{2\cdot j}^\top\wt\bZ_{t-1,\lag}^\top
    \wt\bZ_{t-1,\lag} \bW_{2\cdot j}+ \xi_j+\nu_2\Big)^{-1}\Big(\sum_t \bW_{2\cdot j}^\top\wt\bZ_{t-1,\lag}^\top \Delta_{2jt}^{(m)}\Big),
    \end{align*}
    where
    $\xi_j = (1-T^{-1}){\sum_t}\sum_{i=1}^{N_1}\sum_{k =
      1}^{N_2}W_{2kj}^2Z_{ik(t-1)}\big(Y_{ik(t-1)} -
    Z_{ik(t-1)}\big)$.
  \end{itemize}

  \end{itemize}

  Let $\wt \bLambda = \bLambda^{(M)}$, $\wt  \bGamma =
  \bGamma^{(M)}$, and
$\wt \bE_t=\Z_t-\wt\bLambda\W_1\Z_{t-1}-\Z_{t-1}\W_2\wt\bGamma$.
Once we obtain $\wt\bLambda$ and $\wt\bGamma$, we can proceed to estimate
$\bbeta$ and $\bB$.
Recall that we assume the column space of $\bX$
to be orthogonal to the column space of $\bB$. Therefore
we can estimate $\bbeta$ by
\begin{align}
\wh\bbeta =
  \Big(\bX^\top\bX+\nu_3\bI_p\Big)^{-1}\Big(T^{-1}\sum_t\bX^\top
  \wt\bE_t\Big).\label{wh_beta}
\end{align}
  Subsequently, we obtain $\wh \bB^{(1)}$ by
    $\wh \bB^{(1)} = T^{-1} \sum_t \bP_{\X}^\perp\wt\bE_t$, where
    $\bP_{\X}^\perp=\I-\bP_{\X}$ and $\bP_{\X}=\X(\X\trans\X)^{-1}\X\trans$.
   For any matrix $\bSigma$, let $\bU\bD \bV^\top $ be the SVD decomposition of
  $\bSigma$. Define the soft-thresholding operator $\tau_c$ as
  $\tau_c(\bSigma) = \bU\diag\{(\sigma_i - c)_+ \}\bV^\top$,
  where $\sigma_i=\bD_{ii}$
  is the $i$th diagonal element of $\bD$.
Following \cite{mao2019matrix}, we estimate $\bB$ by
  \begin{align}
  \wh \bB = \frac{1}{1+(1-\alpha)\nu_4}\tau_{\alpha\nu_4/2}(\wh
    \bB^{(1)}).\label{wh_B}
  \end{align}
The above analysis suggests that the solution to (\ref{Q_p2})
has analytical forms,
  which facilitates efficient implementation.
  As shown by \cite{mao2019matrix}, minimizing (\ref{Q_p2}) with
  respect to $\bB$ under the orthogonality constraint to $\bX$ is
  equivalent to the following optimization problem,
  \begin{align*}
  \arg\min_{\bB\in \mR^{N_1\times N_2}}
  \Big\|T^{-1} \sum_t \bP_{\X}^\perp\wt\bE_t - \bB \Big\|_F^2+
\nu_4\Big(\alpha\|\bB\|_*+(1-\alpha) \|\bB\|_F^2\Big).
  \end{align*}
By the Theorem 1 of \cite{mazumder2010spectral}, the analytical
solution is then given by (\ref{wh_B}).
Due to its simple analytical form, we are able to greatly reduce the
computational burden, when compared to other iterative-type algorithms
\citep{troyanskaya2001missing,ma2011fixed}.

\section{Theoretical Properties}\label{sec:theory}

We investigate the theoretical properties of the estimators in this
section.
First, we derive the estimation consistency and normality of
$\wt\blambda$ and $\wt\bgamma$.
In addition, the estimation error bounds for both $\wh \bbeta$ and
$\wh \bB$ are given.
Subsequently, we note that a non-ignorable bias exists for the
first step estimator.
The bias can be large especially for large-scale networks with short time periods.
Therefore we further devise a bias reduction procedure to reduce the
estimation bias.

\subsection{Technical Conditions}

To analyze the theoretical properties of the estimators, the following
technical conditions are required.

\begin{assumption}\label{assum:e}
  {\sc (Distribution)}
  Assume $\ve_{ijt}$ ($1\le i\le N_1, 1\le j\le N_2, 1\le t\le T$) are
  independent and identically distributed sub-Gaussian variables with
  zero mean and a scale factor $0<\sigma<\infty$. That is, we have
  $E\{\exp(t\ve_{ijt})\} \le \exp(\sigma^2t^2/2)$ for any $t$.
  Let
  $E(\ve_{i_1j_1t_1} \ve_{i_2j_2t_2}\ve_{i_3j_3t_3}) =0$
  for any $1\le i_k \le N_1, 1\le j_k\le N_2, 1\le t_k \le T$ with $k = 1,2,3$.
\end{assumption}

\begin{assumption}\label{assum:net_structure}
{\sc (Network Structure)}

\begin{itemize}
  \item [(a)]
  {\sc (Connectivity)}
Treat $\bW_1$ and $\bW_2^\top$ as transition probability matrices of
two Markov chains.
The state spaces are defined as the set of
nodes $\{1, \cdots, N_k\}$ respectively for $k = 1,2$.
We assume the Markov chains are irreducible and aperiodic.
Further define $\bpi_k =
(\pi_{k1},\cdots,\pi_{kN_k})^\top\in\mR^{N_k}$
as the stationary distribution of the $k$th Markov chain,
such that
(i) $\pi_{ki}\geq 0$ and $\sum_{i=1}^{N_k} \pi_{ki} = 1$,
(ii) $\bpi_1= \bW_1^\top\bpi_1$ and $\bpi_2 = \bW_2\bpi_2$.
Assume there exists a finite integer $K$ and a constant $C$ such that $\bW_k^n\cle C \one_{N_k}\bpi_k^\top$ for $n\ge K$ and $k = 1,2$.
  \item [(b)]
  {\sc (Uniformity)}
Assume $\sigma_1(\bW_k) = O (\log N_k)$ for $k = 1,2$.
In addition we assume $\|\bW_1^\top{\one_{N_1}} \|_\infty = O(\log N_1)$
and $\|\bW_2{\one_{N_2}} \|_\infty = O(\log N_2)$.
\end{itemize}

\end{assumption}

\begin{assumption}\label{assum:missing}
  {\sc (Missing Rate)}
  Assume $\min_{i\in [N_1]} p_i \ge c_p$, where  $c_p>0$ and we allow $c_p\to0$ as $m\to\infty$.
\end{assumption}

\begin{assumption}\label{assum:station}
  {\sc (Stationarity)}
  Let $\kappa_1  = \|\bLambda\|_{\max}$,
$\kappa_2  = \|\bGamma\|_{\max}$, and assume
$\kappa_1 + \kappa_2 <\kappa$ with $\kappa<1$.
\end{assumption}

\begin{assumption}\label{assum:bound}
  {\sc (Uniform Boundedness)}
Treat $\X$ as
  fixed covariates and assume
  $|X_{ij}|<C<\infty$ for all $i=1, \dots, N_1$ and $j=1, \dots, p$, where $C$ is a finite constant.
  Let  $\|N_1^{-1}\bX^\top\bX- \bSigma_X\| = o(1)$ as $N_1\rightarrow \infty$,
where $\bSigma_X$ is a positive definite matrix.
  In addition, assume $\|\bbeta_{\cdot j}\|_{2}<C<\infty$ for all $j=1, \dots, N_2$ and $
  \max\{\|\bX\bbeta\|_{\max},\|\bB\|_{\max}\} <C< \infty$.
\end{assumption}

\begin{assumption} \label{assum:iden}
{\sc (Identification)}
  Assume $\bX^\top\bB = \zero$.
\end{assumption}

\begin{assumption}\label{assum:eigs}
{\sc (Local Convexity)}
  Write $\wt Q_p(\btheta, \balpha)$ as a function of $\btheta$ and $\balpha$.
  Let $\ol \bH_{\theta\theta} = E\{\partial^2{\wt Q}_p(\btheta, \balpha)/\partial \btheta\partial \btheta^\top\}$
  and $\ol \bH_{\theta\alpha} = E\{\partial^2{\wt Q}_p(\btheta, \balpha)/\partial \btheta\partial \balpha^\top\}$.
  Assume $\lambda_{\min}(\ol \bH_{\theta\theta}/(mT))\ge \tau_{1}$
  and $\sigma_{1}(\ol \bH_{\theta\alpha}/(mT))\le \tau_2$ as $\min\{m,T\}\rightarrow \infty$,
  where $\tau_1$ and $\tau_2$ are positive constants.
\end{assumption}


\begin{assumption}\label{assum:Sigma_alpha}
 Let  $\bSigma_{\balpha}^{(m)} = (N_1c_p)^{-1}\sum_{i =
   1}^{N_1-1}\bX_i\bX_i^\top p_i(1-p_i)$.
Assume there exists $\bSigma_{\alpha}$ s.t.
$\|\bSigma_{\alpha}^{(m)} - \bSigma_{\alpha}\| = o(1)$
  as $m\rightarrow\infty$ with $c_2\le \lambda_{\min}(\bSigma_{\alpha})\le  \lambda_{\max}(\bSigma_{\alpha})\le c_1$, where
  $c_1$ and $c_2$ are positive constants.
\end{assumption}

Assumption \ref{assum:e} implies that
for a vector $\a = (a_1,\cdots, a_{N_1N_2})^\top$, it holds that
\beq
P\big(\big|\a^\top\E_t\big|>x\big)\le 2\exp\Big(-\frac{x^2}{2\sigma^2\|\a\|^2}\Big) \label{subG_tail}
\eeq
according to \cite{wang2013calibrating},
 where $\E_t = \vec(\mE_t)$.
This condition is widely assumed in high-dimensional modelling
literature \citep{wang2013calibrating,lugosi2019sub,fan2021augmented},
which is more relaxed
than the normality assumption.
Next, Assumption \ref{assum:net_structure} is a condition imposed on
the network structure.
Assumption \ref{assum:net_structure}(a) requires that all the network
nodes are reachable by each other (i.e., irreducibility). A
simple and sufficient condition for both irreducibility and
aperiodicity is that the network is always fully connected after a
finite number of steps.
According to the famous small world property of the networks,
this
condition can be easily satisfied by a  variety of real world networks
\cite[Chapter~4.2]{watts1998collective,newman2011structure}.
Next, we assume that $\bW_k^n$ is elementwisely bounded by
$C\one_{N_k} \bpi_k^\top$ for $n\ge K$, which is a direct conclusion if
the Markov chain converges to its stationary distribution uniformly
 at a fast rate.
The condition can be relaxed to allow $K$ slowly diverge to infinity
as $m\to \infty$.
Assumption \ref{assum:net_structure}(b) is a uniformity condition,
which restricts the  superstar effect in the network
\citep{zhou2017estimating,zhu2021network}.

Subsequently, Assumption \ref{assum:missing} is about the
non-missing rate, which is allowed to go to zero
as $N_1$ and $N_2$ diverge.
Consequently, we allow the scenario of sparse observation entries.
Similar setting is also considered by \cite{mao2019matrix} for static data.
In addition, the specification of the missing mechanism is more
flexible for matrix autoregression model considered by
\cite{chen2021autoregressive}, which does not allow missingness in
the matrix entries.

Assumption \ref{assum:station} is imposed to assure the
  stationarity of the matrix-valued time series.
 Here the stationarity of the matrix-valued time series refers to the
 stationarity of the corresponding vector formed time series, i.e.,
 $\vec(\bY_t)$.
  Similar conditions are imposed by \cite{zhu2017network} for dynamic
  network data.
Assumption \ref{assum:bound} treats $\bX$ as bounded fixed covariates
for theoretical convenience.
Assumption \ref{assum:iden} is an identification condition, which
allows us to separately estimate $\bbeta$ and $\bB$
\citep{mao2019matrix}.
Next, Assumption \ref{assum:eigs} assumes local convexity of the
objective function.
Assumption \ref{assum:Sigma_alpha} assumes that
we can obtain a reliable estimation of $\balpha$ in (\ref{logistic_R})
by using the maximum likelihood estimation with the log-likelihood
function as
$\mL(\balpha) = \sum_{i,j,t}\{R_{ijt} \bX_i^\top\balpha -
  \log(1+\exp(\bX_i^\top\balpha))
  \}$ and $\bSig_\balpha^{(m)}$ is the Hessian matrix.
  The assumption uses $c_p$ here to limit the effect of the
  unbalanced data (i.e., $P(R_{ijt} = 1)\to 0$). Similar assumption has
  been used in recent literature about unbalanced data learning
  problems \citep{wang2020logistic}.

\subsection{First Step Estimation Properties}

Define $\wt\btheta = \arg\min_{\btheta} \wt Q_p(\btheta)$, where $\wt
Q_p(\btheta)$ is given in (\ref{Q_obj}).
We first establish the asymptotic properties of the profile estimator
$\wt\btheta$ in the following theorem.

\bet\label{th:normal}
 Assume Assumptions \ref{assum:e}--\ref{assum:Sigma_alpha} hold.
Let $\bfeta$ be an arbitrary vector in $\mR^m$ with $ c_1\le
\|\bfeta\|\le \|\bfeta\|_1 \le c_2$, where $c_1,c_2$ are
  positive constants.
Recall that $\wt \bSigma_2 = (mT)^{-1}\ddot{\wt Q_p}(\btheta_0)$
and $\bSigma_2 = E(\wt \bSigma_2)$.
The explicit form of $\wt \bSigma_2$ is given in Section A.2 in the supplementary materials.
In addition,
let $(mT)^{-1}c_p^2\bfeta^\top \cov\{\dot{\wt
  Q_p}(\btheta_0)\}\bfeta\rightarrow \bfeta^\top\bSigma_1\bfeta$,
where $\bSigma_1$ is a positive definite matrix.
The specific form of $\bSigma_1$ is also given in Section A.2
in the supplementary materials.
Further assume that
\begin{align}
\frac{1}{Tc_p^4}\Big[&mr_{\pi}^2 (\log m)^{4K+1}+mr_{\pi}(\log m)^{2K+3}+ (\log m)^{2K}+
c_p^2 m^2r_\pi^2 (\log m)^{4K}\nonumber\\
&+c_p^3\big\{
m^4r_\pi^4 (\log m)^4 + m^2r_\pi^2 (\log m)^{4K+4}\big\}\Big]\rightarrow 0,\label{NT_cond}
\end{align}
where  $r_\pi = \max(r_{1\pi}, r_{2\pi})$,
$r_{1\pi} = \bpi_1^\top\bpi_1 $, $r_{2\pi} = \bpi_2^\top\bpi_2$,
and $K$ is given in Assumption \ref{assum:net_structure}(a).
Define
$ \b^{(1)} = ( \b^{(1)\lambda\top},  \b^{(1)\gamma\top})^\top$ with
  $ \b^{(1)\lambda} = ( b_i^{(1)\lambda}:1\le i\le N_1)^\top$ and
  $ \b^{(1)\gamma} = ( b_j^{(1)\gamma}:1\le j\le N_2)^\top$,
  i.e.,
  \begin{align}
  &b_i^{(1)\lambda} = \frac{2}{mT}\lambda_i\sum_j \sum_k W_{1ik}^2E\Big\{Z_{kjt}(Y_{kjt} - Z_{kjt})\Big\},\label{bias_lambda1}\\
  &b_j^{(1)\gamma} = \frac{2}{mT}\gamma_j\sum_i \sum_k W_{2kj}^2E\Big\{Z_{ikt}(Y_{ikt} - Z_{ikt})\Big\}.\label{bias_gamma1}
  \end{align}
Then we have
\begin{align}
\sqrt{mT}c_p\bfeta^\top (\wt\btheta - \btheta - \b_1) \rightarrow_d N(0,
  \bfeta^\top\bSigma_2^{-1}\bSigma_1\bSigma_2^{-1}\bfeta),\label{wt_theta_normal}
\end{align}
where $\b_1 = \wt\bSigma_2^{-1}(\b^{(1)}- \br/(mT))$
and $\br = 2(\nu_1 \blambda^\top , \nu_2\bgamma^\top)^\top$.
 Further assume $\max\{\nu_1, \nu_2\} = o(c_p^{-1}m (\log m)^{2K})$,
 then  we have
$\|\b^{(1)}\|_\infty= O(c_p^{-1} (\log m)^{2K}/T)$
and $ \|\b_1\|_\infty=  O_p(c_p^{-1}(\log m)^{2K}/T)$.
\eet

Regarding the theoretical results, we have the following comments.
First, the condition in (\ref{NT_cond}) is a critical condition, which requires $T$
to be sufficiently large to obtain a reliable estimation of $\btheta$.
This condition is easier to hold if $m$ (the network sizes $N_1+N_2$) and
$r_\pi$ (the quantity related to the stationary distributions of $\bW_1$ and $\bW_2$) are lower, and $c_p$ (the lower bound of observation rate) is higher.
Recall that $\bpi_1$ and $\bpi_2$ are stationary distribution vectors
corresponding to $\bW_1$ and $\bW_2$, respectively.
Take $\bpi_1$ as an example.
If the stationary distribution is relatively uniform, we should have
$\pi_{1i}\approx N_1^{-1}$ ($1\le i\le N_1$), which leads to
$\bpi_1^\top \bpi_1 \approx N_1^{-1}\to 0$.
In this case, $r_{1\pi}$ converges to zero at a fast speed.
In the network literature, $\bpi_1$ and $\bpi_2$ are also referred to
as {\it eigenvector centrality} of the network nodes
\citep{jackson2010social}.
The eigenvector centrality is typically used to characterize the
influential power of the network nodes.
If the distribution of the influential power of the network nodes is
relatively uniform, we will have $\pi_{1i}\approx N_1^{-1}$  and
$\pi_{2i}\approx N_2^{-1}$. In other words, $r_\pi$ will converge to
zero at a fast rate if
the network superstar effect is low.
In the meanwhile, the term $\log(m)$ is also closely related to the
uniformity Assumption \ref{assum:net_structure}(b).
In Assumption \ref{assum:net_structure}(b), we basically assume
$\sigma_1(\bW_k) = O(\log m)$ and $\|\bW_1^\top\one\|_\infty =O(\log m)$.
Consider an extreme case for instance. Let $a_{1, i1} = 1$ for $2\le
i\le N_1$ and the other entries in $\bA_1$ be zero.
In this case, the first node is a superstar in the network and one can
verify that $\sigma_1(\bW_1) = O(m^{1/2})$ and $\|\bW_1^\top \one\|_\infty = O(m)$.
This extreme case breaks the uniformity of the nodes and
violates  Assumption \ref{assum:net_structure}(b).
As a consequence,
the condition in (\ref{NT_cond}) can be satisfied if the network
superstar effect and missing rate are both controlled.

Second,
the asymptotic convergence result in (\ref{wt_theta_normal}) implies
that the asymptotic bias is given by $\b_1$ and the convergence rate
is given by $\sqrt{mT}c_p$.
The asymptotic bias is mainly determined by the time length
$T$, the  network structure related quantity $(\log m)^{2K}$,
and the observation rate $c_p$.
Consider an elementwise convergence by setting $\bfeta = \e_i$ for $1\leq i \leq m$.
The asymptotic bias term will
  vanish as long as
$T^{-1}m(\log m)^{4K}\to 0$,
which requires relatively larger number of
time periods $T$.
 Note that the convergence rate
 increases when $m$ or $T$ increases. The increment with $T$ is
 directly due to the increased
 number of time periods. The increment with $m$ can be understood as
a result of the
 cumulated information contributed by the matrix products
 $\W_1\bY_{t-1}$ and $\bY_{t-1}\W_2$, which correspond to
combining $N_1$ and $N_2$ observations.
Next, the convergence rate will be faster if $c_p$ is higher.
This is because a higher observation rate $c_p$  will result in a
larger effective sample size, and thus better estimation performance.
The asymptotic result extends that of the classical panel data model
with fixed effects \citep{arellano2007understanding}  and our focus is on
 matrix-valued time series with incomplete
observations.
Lastly, to facilitate the diverging dimension of the parameter $\btheta$,
we describe the asymptotic normality
in (\ref{wt_theta_normal}) through its projection to
an arbitrary vector $\bfeta$  satisfying $ c_1\le \|\bfeta\|\le \|\bfeta\|_1 \le c_2$.
The reason that allows us to consider an arbitrary $\bfeta$ with
bounded and positive
$L_1$ and $L_2$ norms is that, our parameter dimension is not ultra-high, although it is diverging.
In this case, the projection of the estimated parameter to an
arbitrary direction remains asymptotically normal. This is no longer
the case if we consider ultra high-dimensional parameters.
For statistical inference in ultra high dimensions, we refer to
\cite{chernozhukov2017central}, \cite{chang2021central}, and
\cite{koike2022high} for
novel theoretical tools for establishing the large sample properties.

\begin{corollary}\label{cor:consistency1}
Assume the same conditions as in Theorem \ref{th:normal}.
Write $\b_1 = (\b_1^{\lambda\top}, \b_1^{\gamma\top})^\top$,
where $\b_1^{\lambda}\in \mR^{N_1}$ and
$\b_1^{\gamma}\in \mR^{N_2}$. Then we have
\begin{align*}
&  N_1^{-1/2}\big\|\wt \blambda - \blambda_0 - \b_1^\lambda\big\|  = O_p((mT)^{-1/2}
c_p^{-1}), \\
&  N_2^{-1/2}\big\|\wt \bgamma - \bgamma_0- \b_1^\gamma\big\|  =
  O_p\big((mT)^{-1/2}c_p^{-1}\big).
\end{align*}
\end{corollary}
\noindent
Corollary \ref{cor:consistency1} is a direct conclusion of
Theorem \ref{th:normal}.
It establishes the $L_2$-convergence result of the parameter
$\wt\btheta$.
Given the convergence result of $\wt\btheta$, we are able to
establish the estimation error bounds for $\wh \bbeta$
and $\wh \bB$ respectively in the following theorem.

\bet\label{th:beta_bound}
Assume the same conditions as in Theorem \ref{th:normal} and $
\nu_3=  o(m)$, then we have
\begin{align}
\frac{1}{N_2p} \|\wh\bbeta - \bbeta\|_F = O_p\Big\{\frac{1}{N_2p}\Big(
m^{-1/2}\nu_3
+ (Tc_p)^{-1/2}p^{1/2}(\log m)^K + p^{1/2}(\|\b_1^\lambda\| + \|\b_1^\gamma\|)\Big)\Big\}.\label{beta_diff}
\end{align}
\eet

\noindent
Theorem \ref{th:beta_bound} establishes the error bound for
$\wh \bbeta$.
Here $N_2\times p$ is the dimension of $\bbeta$ matrix.
Therefore $\|\wh\bbeta - \bbeta\|_F/(N_2p)$ is the average error bound
for $\wh \bbeta$ matrix.
As shown in (\ref{beta_diff}), the error of $\wh \bbeta$ constitutes of three major terms.
The first term $(m^{-1/2}\nu_3)/(N_2p)$ is contributed by the estimation bias
caused by using ridge regression with penalty factor $\nu_3$.
Set $\nu_3 = o(m)$, then we have $(m^{-1/2}\nu_3)/(N_2p) = o(m^{-1/2}p^{-1})$.
The remaining two terms are caused by the estimation of
$\btheta$  in our first step.
Specifically, the second term $\{(Tc_p)^{-1/2}(\log m)^K\}/(N_2p^{1/2})$ captures the effect from
the estimation variability of $\btheta$, while the third term
$(\|\b_1^\lambda\| + \|\b_1^\gamma\|)/(N_2p^{1/2})$ captures the effect from the
estimation bias of $\btheta$. It is notable that under (\ref{NT_cond}), the second term dominates the third term. However, we still keep them both in (\ref{beta_diff}) to reflect the two effects explicitly.
Obviously, increasing $T$ will help to  reduce the second and third
terms, but not the first term.
Similar to our previous results, the network  structure related
quantity $(\log
m)^K$ and the observation rate $c_p$ play important roles in the final
error bounds.

\bet\label{thm:B_bound}
Assume the same conditions as in Theorem \ref{th:normal}.
In addition let $ \nu_4'
  \alpha\Delta_{E,N_1N_2T}^{-1}\to\infty$
and $p\ll
  (c_pN_1N_2T)^{1/2}$,
where $\nu_4' =\nu_4/(N_1N_2)$, and
\begin{align*}
\Delta_{E,N_1N_2T}  &=
\frac{1}{m^2}\Big\{\frac{p+m^{1/2}\log(m)}{T^{1/2}c_p}
+\frac{\sqrt{m\log(m)}\{1
+\log(T) r_\pi^{1/2} (\log m)^{K+1}\}}{\sqrt{c_p T}}
+ m\|\b_1\|_\infty\Big\}.
\end{align*}
Then we have
\begin{align}
\frac{1}{N_1N_2}\|\wh\bB - \bB\|_F^2 = O_p\Big(\max[\min\{
\nu_4'\alpha\|\bB\|_*,
N_2N_2r_B(\nu_4'\alpha)^2\},
\nu_4'(1-\alpha)\|\bB\|_F^2
]\Big),\label{B_diff}
\end{align}
where $r_B$ is the rank of $\bB$.
\eet
\noindent
Theorem \ref{thm:B_bound} indicates that the error bound for $\bB$ is
related to the low rank structure of $\bB$
(i.e., $\|\bB\|_*$ and $r_B$), the penalty factors (i.e., $\nu_4$ and $\alpha$),
and the Frobenius norm of $\bB$ (i.e., $\|\bB\|_F^2$).
The general proof follows the guideline of \cite{mao2019matrix}
and the conclusion (\ref{B_diff}) is also consistent
with Theorem 1 of \cite{mao2019matrix}.
The main difference is the formulation of $\Delta_{E,N_1N_2T}$, which
is actually the error bound for $(N_1N_2)^{-1}\|\bX\bbeta + \bB -
\ol \bE\|$, where $\ol \bE = T^{-1}\sum_t\wt \bE_t$.
Establishing this error bound is more challenging in our
case since we need to further deal with the time dependence
of matrix-valued time series.
To this end, we use the tools of martingale difference arrays for matrix-valued data \citep{tropp2011freedman}, as well as the random matrix theories
\citep{ajanki2017universality,Erods18,erdHos2019bounds} to establish the error bound.
We remark that the second and third terms in the expression of
$\Delta_{E,N_1N_2T}$ are induced by the estimation error
of $\wt\btheta$ in the first step.
Particularly, compared to the static matrix completion result of \cite{mao2019matrix}, the error rate can be further reduced by
increasing $T$ in our case.

Lastly, we note that the estimation bias $\b_1$ influences the
asymptotic property of $\wt\btheta$. It
is also linked to $\|\wh \bbeta - \bbeta\|$ in Theorem \ref{th:beta_bound}
and $\|\wh\bB - \bB\|_F^2$ in Theorem \ref{thm:B_bound}.
Then it is of great interest to study how to reduce the estimation
bias to obtain better estimators.

\subsection{Bias Reduction and Correction}

 In this section we study how to reduce the estimation bias
in the first step estimation, which  enables us to further improve
 the estimation efficiency for
$\wh \bbeta$ and $\wh \bB$.
Specifically, we can estimate the bias terms in (\ref{bias_lambda1})
and (\ref{bias_gamma1}) using the matrix data through
\begin{align}
  &\wh b_i^{(1)\lambda} =  \frac{2}{mT}\wt\lambda_i\sum_j \sum_k
    W_{1ik}^2\frac{1}{T}\sum_t\Big\{\wh Z_{kjt}^2(\wh  p_k - 1)\Big\},\label{bias_lambda2}\\
  &\wh b_j^{(1)\gamma}= \frac{2}{mT}\wt\gamma_j\sum_i \sum_k
    W_{2kj}^2\frac{1}{T}\sum_t\Big\{\wh Z_{ikt}^2(\wh p_i -1)\Big\},\label{bias_gamma2}
  \end{align}
 where
 $\wh Z_{kjt} = Y_{kjt}R_{kjt}/\wh p_k$
  and $\wh \b^{(1)} = (\wh \b^{(1)\lambda\top}, \wh \b^{(1)\gamma\top})^\top$.
  Define the debiased estimator as
  $
  \wh\btheta = \wt\btheta - \wh \b_1
  $,
 where $\wh \b_1 = \wt\bSigma_2^{-1}( \wh \b^ {(1)}-(mT)^{-1}\wh \r)$
 and $\wh\r=2(\nu_1\wt\blambda^\top, \nu_2\wt\bgamma^\top)^\top$.
Here $\wh \b_1$ is the estimator for $\b_1$ defined in (\ref{wt_theta_normal}).
The following theorem establishes the asymptotic properties
for the debiased estimator $\wh \btheta$.

\bet\label{th:bias_reduce}
{\sc(Bias Reduction)}
Assume the same conditions as in Theorem \ref{th:normal}.
Further assume $ T\gg \max\{
  c_p^{-1/2}p(\log m)^{2K},(\nu_1+\nu_2)^{2/3}(\log m
  )^{4K/3}\}$.
Define
\begin{align*}
& b_{i}^{(2)\lambda} = \frac{1}{mT}\sum_k W_{1ik}^2\sum_j  E\{Z_{kjt}(Y_{kjt} - Z_{kjt})\}b_{1i}^\lambda,\\
& b_{j}^{(2)\gamma}= \frac{1}{mT}\sum_k W_{2kj}^2\sum_i  E\{Z_{ikt}(Y_{ikt} - Z_{ikt})\} b_{1j}^\gamma,
\end{align*}
and $\b^{(2)} = ( \b^{(2)\lambda\top}, \b^{(2)\gamma\top})^\top$ with
  $\b^{(2)\lambda} = (b_{i}^{(2)\lambda}:1\le i\le N_1)^\top$ and
  $\b^{(2)\gamma} = ( b_{i}^{(2)\gamma}:1\le i\le N_2)^\top$.
  Further define  $\b_2 = -\wt\bSigma_2^{-1}\b^{(2)} $.
   Then we have
  \begin{align}
  \sqrt{mT}c_p\bfeta^\top \wt \bSigma_2 \big(\wh \btheta - \btheta - \b_2\big)
  \rightarrow N(0,\bfeta^\top \bSigma_1\bfeta).\label{normal_debias}
  \end{align}
   In addition we have $\|\b_2\|_\infty =  O_p(T^{-2}
   (\log m)^{4K}c_p^{-2})$.
\eet
\noindent
Theorem \ref{th:bias_reduce} implies that,
the asymptotic bias  for the debiased estimator $\wh \btheta$ is given
by $\b_2$ and
the convergence rate remains $\sqrt{mT}c_p$, which is the same as in
Theorem \ref{th:normal}.
In particular, we note that $\|\b_2\|_\infty = \|\b_1\|_\infty O_p(
c_p^{-1}T^{-1}(\log m)^{2K}) = O_p(\|\b_1\|_\infty^2)$.
Condition (\ref{NT_cond}) implies that $c_p^{-1}T^{-1}(\log
m)^{2K}\to0$. Therefore the bias $\b_2$ is further reduced compared to $\b_1$.
Moreover, motivated by the result above,
we can apply the bias reduction procedure multiple times to further reduce the bias.
Suppose in the $r$th step ($r\ge 2$),
the bias estimator is  $\wh \b_r = (\wh \b_r^{\lambda\top}, \wh \b_r^{\gamma\top})^\top$.
We have $\wh \b_{r} = (-1)^{r+1}\wt \bSigma_2^{-1} \wh \b^{(r)}$ with
$\wh \b^{(r)} = (\wh \b^{(r)\lambda\top },\wh \b^{(r)\gamma\top })^\top$,
where
\be\label{eq:estbiasupdate}
&&\wh b_{i}^{(r)\lambda} = \frac{1}{mT^2}\sum_k W_{1ik}^2\sum_j \sum_t
\{\wh Z_{kjt}^2(\wh  p_k -1)\}\wh b_{r-1,i}^\lambda\n,\\
&&\wh b_{j}^{(r)\gamma} = \frac{1}{mT^2}\sum_k W_{2kj}^2\sum_i  \sum_t
\{\wh Z_{ikt}^2(\wh p_i - 1)\}\wh b_{r-1,j}^\gamma.
\ee
Denoting the debiased estimator in the $r$th step as $\wh \btheta_r$ ($r\ge 1$), we have
$\wh\btheta_{r+1} = \wh \btheta_r - \wh \b_{r+1}$, where $\wh \btheta_1$ is set as $\wh \btheta$.
With the number of debias rounds $r$ large enough,
one should be able to obtain a bias corrected estimator.
We state this result in the following theorem.

\bet\label{thm:bias_corrections}
  {\sc (Bias Correction)}
  Assume the same conditions as in Theorem \ref{th:normal}.
For integers $r=1, 2, \dots$,
  define $\b_r = (\b_r^{\lambda\top}, \b_r^{\gamma\top})^\top$ and
  $ \b_{r} = (-1)^{r+1}\wt \bSigma_2^{-1}\b^{(r)}$ with
  $ \b^{(r)} = ( \b^{(r)\lambda\top }, \b^{(r)\gamma\top })^\top$, where
  \begin{align}
& b_{i}^{(r)\lambda} = \frac{1}{mT}\sum_k W_{1ik}^2\sum_j  E\{Z_{kjt}(Y_{kjt} - Z_{kjt})\} b_{r-1,i}^\lambda,\label{b_i_r1}\\
& b_{j}^{(r)\gamma} = \frac{1}{mT}\sum_k W_{2kj}^2\sum_i   E\{Z_{ikt}(Y_{ikt} - Z_{ikt})\} b_{r-1,i}^\gamma.\label{b_j_r1}
\end{align}
Then $\|\b_{r+1}\|_\infty=\|\b_r\|_\infty O(c_p^{-1}(\log m)^{2K}/T)$.
Let $\wh\btheta_r=\wh\btheta_{r-1}-\wh\b_r$, where $\wh\b_r$ is
  given in (\ref{eq:estbiasupdate}).
Assume $\Delta_{N_1N_2}=o(T)$.
Then we have
\beq
\sqrt{mT}c_p\bfeta^\top\wt \bSigma_2\Big\{\wh \btheta_r - \btheta - \b_{r+1}
\Big\}\rightarrow_d N(0,
  \bfeta^\top\bSigma_2^{-1}\bSigma_1\bSigma_2^{-1}\bfeta).\label{theta_r_normal}
\eeq
Furthermore, we have
\beq
\sqrt{mT}c_p\bfeta^\top(\wh \btheta_r - \btheta)\rightarrow_d N(0,
  \bfeta^\top\bSigma_2^{-1}\bSigma_1\bSigma_2^{-1}\bfeta)\label{bias_correct_normal}
\eeq
for  $r> \log\{m(\log m)^{2K}/T^{1/2}\}/\log[T
c_p/\{C(\log m)^{2K}\}]$, where
$\|\b^{(1)}\|_\infty \lesssim C (\log m)^{2K}/(Tc_p)$.
\eet
\noindent
Theorem \ref{thm:bias_corrections} states that
the asymptotic bias
can be ignorable when sufficient number of debiasing rounds
are conducted.
As indicated by the result,
let $T = O(m^\delta)$ for $\delta>0$ and $c_p\ge c_0$ for a positive
constant $c_0$, then only a finite number of rounds is needed
for correcting the bias.
In our numerical studies, we find that typically 2--3 rounds of bias
reduction will be sufficient to obtain a reliable estimator.

\section{Numerical Studies}\label{sec:numerical}

\subsection{Simulation Design}

To demonstrate the finite sample performance of the MNAR model, we present in this section a variety of simulation experiments. We first discuss how to generate networks $\bA_1\in \mR^{N_1\times N_1}$ and $\bA_2\in \mR^{N_2\times N_2}$.
We refer to the two networks as row-network and column-network respectively in the following.
Throughout the simulation study, we assume both the row-network and column-network follow the power-law distribution model.
That is, we consider the power-law distribution of in-degrees of network nodes, which reflects a popular network phenomenon in practice. It implies that, only a small amount of nodes in the network have a large number of followers (i.e., in-degrees),
while the majority of nodes have very few followers.
We follow \cite{clauset2009power} to generate the row and column
adjacency matrices $\bA_1 \in \mR^{N_1\times N_1}$ and $\bA_2 \in
\mR^{N_1\times N_1}$  with the power-law distribution pattern.
Specifically, for each node $i$ ($1\leq i \leq N_1$), we first generate its in-degree $d_{1i}=\sum_ja_{1,ji}$ according to the discrete power-law distribution, i.e., $P(d_{1i}=h)=ch^{-\upsilon}$, where $c$ is a normalizing constant and $\upsilon$ is the exponent parameter.
Then, we randomly assign $d_{1i}$ nodes to be the followers of node $i$. The column-network  $\bA_2$ is generated similarly.
In both networks, we set $\upsilon=2.5$. After generating $\bA_1$ and
$\bA_2$, the row-normalized adjacency matrix $\bW_1$ and the
column-normalized adjacency matrix $\bW_2$ are calculated accordingly.

Given the network structures, we proceed to generate the matrix-valued time series $\{\bY_t:1\le t\le T\}$.
First, let $p=6$ and denote the covariate matrix as $\bX=(\bm{1}_{N_1},\bX_{\text{sub}}) \in \mR^{N_1\times p}$, where $\bX_{\text{sub}}$ is generated from a standard normal distribution.
Subsequently, we generate the corresponding coefficient matrix $\bbeta$ as follows. We first generate a $p\times N_2$ dimensional matrix with each element independently following $U(-0.01,0.01)$.
Then we randomly set 95\% elements to be zero, which yields a sparse
coefficient matrix $\bbeta$.
To generate  the low-rank matrix $\bB$, we first set $r=10$ and generate $\bU \in \mR^{N_1\times r}$ and $\bV \in \mR^{N_2\times r}$ from a normal distribution with mean zero and standard deviation  0.5.
Then, we compute $\bB=\bP_{\X}^\perp\bU\bV^{\top}$, which ensures the
column spaces of $\bB$ and $\bX$ to be orthogonal to each other.
Next, for the diagonal matrices $\bLambda$ and $\bGamma$, we set $\lambda_i=0.45$ for $1\leq i \leq N_1$ and $\gamma_i=0.45$ for $1\leq i \leq N_2$.
Finally, the entries in the noise matrix $\mE_t \in \mR^{N_1\times
  N_2}$ with $1 \leq t \leq T$ are independently generated from the
standard normal distribution. Having generated $\bLambda$, $\bGamma$, $\bX$,
$\bm{\beta}$, $\bB$, and $\mE_t$, we generate the matrix $\bY_t$ according to the MNAR model \eqref{model}.

After obtaining $\bY_t$, we follow \cite{mao2019matrix} to consider two missing mechanisms for generating the observed matrix $\bZ_t$.
The first missing mechanism considered here is \emph{missing at
  random} (MAR), in which we adopt the logistic regression model to
generate the observation indicator $R_{ijt} = 1$ for $1\leq i \leq
N_1$, $1\leq j \leq N_2$, and $1\leq t \leq T$. Specifically, assume
the coefficient vector
$\bm{\alpha}=(\alpha_0,\alpha_1,...,\alpha_p)^{\top} \in \mR^{p+1}$,
and set $\alpha_0=-1.3$, $\alpha_i=0.1$ for $1\leq i \leq p$. Then the
observation probability $p_{i}$ is computed as $p_{i} =
\exp(\bX_{i}^\top \balpha)/\{1+\exp(\bX_{i}^\top \balpha)\}$. The
observation indicator $R_{ijt}$ is then generated from a Bernoulli
distribution with parameter $p_{i}$. The second missing mechanism is
\emph{uniform missing} (UNI), in which we assume all observations have
the same missing probability. To this end, we set the observation
probability as $p_{i}=0.2$ for $1\leq i \leq N_1$. The observation
probability 0.2 is chosen to approximate the average observation
probability under the missing mechanism of MAR.

We set $N_1 = N_2 = N$ and consider $N= (100, 200, 400, 600)$. As for the time span, we consider $T=(30, 60, 100)$. Therefore, it results in a total of $4 \times 3 = 12$ experimental settings under each missing mechanism. In each experimental setting, we repeat the experiment  $R = 200$ times to obtain a reliable evaluation.

\subsection{Performance Measurement}

To evaluate the estimation performance of the MNAR model, we compare
it with the singular value soft-thresholding (SVT) method proposed by
\cite{mao2019matrix}, as it has shown advantages over several
benchmark matrix completion methods.
Since the SVT method is a static approach and cannot handle dynamic
matrix completion problem directly,
we have modified the original SVT method for comparison
purpose. Specifically, we consider three modified versions of the SVT
method. First, the SVT method is applied to a data set at each single
time point,
and the resulting estimators in different times are regarded as
competitors to MNAR.
We call this method SVT-SEP for convenience.
Second, we take average of the SVT estimators at different times to
obtain a more stable estimator, which is called the SVT-AVG
estimator. Lastly, we ignore the time dependence and estimate the
parameters by minimizing the following objective function
\beq
Q_p^{(3)}(\bbeta,\bB)\defeq \frac{1}{T}\sum_{t = 1}^T\big\|\bY_t - \bX\bbeta - \bB\big\|_F^2 +
\nu_3\|\bbeta\|_F^2+
\nu_4\Big(\alpha\|\bB\|_*+(1-\alpha) \|\bB\|_F^2\Big)\nonumber
\eeq
The corresponding estimator is referred to as SVT-SUM.

To estimate the MNAR model and three SVT-type methods, the observation
probabilities $p_i$s need to be estimated first.
Under the missing mechanism of UNI, the averaged value of $\bR_t$ is
used as the estimated observation probability.
Under MAR, a logistic regression model is
first conducted to estimate $\bm{\alpha}$. With the resulting
estimator $\widehat{\balpha}$, the observation probabilities are
calculated as $\widehat{p}_{i} = \exp(\bX_{i}^\top
\widehat{\balpha})/\{1+\exp(\bX_{i}^\top \widehat{\balpha})\}$. To
implement the MNAR method, as we mentioned before, the original
estimators $\wt \bLambda$ and $\wt \bGamma$ are biased. Therefore, we
conduct bias correction to obtain more precise estimators, which are
denoted by MNAR-ADJ in subsequent analysis. Specifically, for
short-term data sets (i.e., $T=30$), we apply the bias reduction
procedure two times; while for long-term data sets (i.e., $T=60,100$),
only one time bias reduction is conducted. We also obtain the original
estimators without using bias reduction for comparison purpose, which
we refer to as MNAR-ORG.

For each method (i.e., MNAR-ORG, MNAR-ADJ, SVT-SEP, SVT-AVG, and SVT-SUM), we define $\widehat{\bbeta}^{(r)}=(\widehat{\beta}_{ij}^{(r)}) \in \mR^{p \times N_2}$ as the estimator for $\bbeta$ in the $r$th replication ($1\leq r \leq R$).
Then, to evaluate the estimation efficiency of each method, we define the root mean squared error (RMSE) for $\bbeta$, namely, $\text{RMSE}_{\bbeta}=\{\sum_{r=1}^{R}\sum_{i,j}(\widehat{\beta}_{ij}^{(r)}-\beta_{ij})^2/(RpN_2)\}^{-1/2}$. The estimation performance of other parameters (i.e., $\bLambda$, $\bGamma$, $\bB$) are calculated similarly.

Next, we proceed to evaluate the performance of the matrix completion task.
Define $\bY_t=\bA_t+\mE_t$, where $\bA_t = E(\bY_t|\mF_{t-1})$ and
$\mF_{t-1} = \{\bY_{s}, \bR_s:s\le t-1\}$.
Subsequently, we evaluate the estimation performance for $\bA_t$, which is the target matrix in matrix completion problems.
For the SVT-type methods, the estimator for $\bA_t$ can be computed directly.
While in the MNAR model, we have $\bA_t=\bLambda \bW_1\bY_{t-1} + \bY_{t-1}\bW_2\bGamma + \bX\bbeta  + \bB$.
The estimation of $\bA_t$ requires recovering the whole matrix $\bY_{t-1}$ first.
To this end, we propose a rolling recovering strategy for $\bY_t$.
Specifically, let $\bZ_{0} = \bZ_{1}$ and first calculate $\widehat{\bA}_1=\widehat{\bLambda} \bW_1\bZ_{0} + \bZ_{0}\bW_2\widehat{\bGamma} + \bX\widehat{\bbeta}  + \widehat{\bB}$,
where we substitute $\{\bLambda, \bGamma, \bbeta, \bB\}$ with their estimates accordingly.
For $1 < t \leq T$, we conduct rolling prediction of $\bA_t$ as
$\widehat{\bA}_t=\widehat{\bLambda} \bW_1\widehat{\bA}_{t-1} + \widehat{\bA}_{t-1}\bW_2\widehat{\bGamma} + \bX\widehat{\bbeta}  + \widehat{\bB}$.
The RMSE of the estimation for the matrix $\bA_t$ is calculated as
RMSE$_\bA = \{(RTN_1N_2)^{-1}\sum_{r = 1}^R \sum_{t=1}^T\|\wh \bA_t^{(r)} - \bA_t\|_F^2\}^{1/2}$,
where $\wh \bA_t^{(r)}$ is the estimation for $\bA_t$ in the $r$th simulation round.
Last, to purely quantify the performance of matrix completion, we consider the measure of test error \citep{mao2019matrix}, which focuses on the missing values in the matrix. The test error is calculated by $\sum_{i,j,t}\{R_{ijt}^\dag(\widehat{A}_{ijt}-Y_{ijt})\}^2/\sum_{i,j,t}(R_{ijt}^\dag Y_{ijt})^2$, where $R_{ijt}^\dag = 1-R_{ijt}$.

\subsection{Simulation Results}

Tables \ref{t:case1} and \ref{t:case2} present the simulation results
under the missing mechanisms of MAR and UNI, respectively. In general,
the simulation results under different missing mechanisms are
similar. Specifically, we can draw the following conclusions. First,
compared with the SVT-type methods, the proposed MNAR-ORG and MNAR-ADJ
estimators have achieved lower RMSEs and lower test errors in all
experimental settings. These results suggest better estimation
performance and matrix completion performance by considering network
information in the MNAR method. Among the SVT-type estimators, in
general, SVT-AVG performs the best, which is followed by SVT-SUM and
SVT-SEP.
Second, by conducting bias reduction, the MNAR-ADJ estimators have
better estimation performance than MNAR-ORG estimators.
This finding demonstrates the usefulness of bias reduction operation.
Particularly, the RMSE$_{\bA}$ and the test error for the MNAR-ADJ
method is lower than the other competing methods, which illustrates
the potential power of the proposed MNAR method in dealing with the matrix completion task.
Last, as the sample size $(N_1,N_2)$ or the time span $T$ increases,
the RMSEs of all estimates in MNAR-ORG and MNAR-ADJ decrease, implying
consistency of the corresponding estimators.

\section{Real Data Analysis}\label{sec:real_data}

To demonstrate the practical performance of MNAR, we
conduct an empirical study using a large public data set on Yelp, which
is the largest review site in the United States. The original data set
is available from \url{https://www.yelp.com/dataset}. This data set
contains detailed information about Yelp's businesses, reviews, and
users during the year 2010 to 2018. The whole data set contains five parts. They are, respectively:
(1) shop information (as restaurants, home services, etc.), (2) user
information, (3) reviews commenting on shops from users, (4) short
tips for shops from users, and (5) the aggregated check-ins of businesses.

\begin{figure}[h]
	\centering
	\subfloat[Las Vegas]{
		\includegraphics[width=0.33\textwidth]{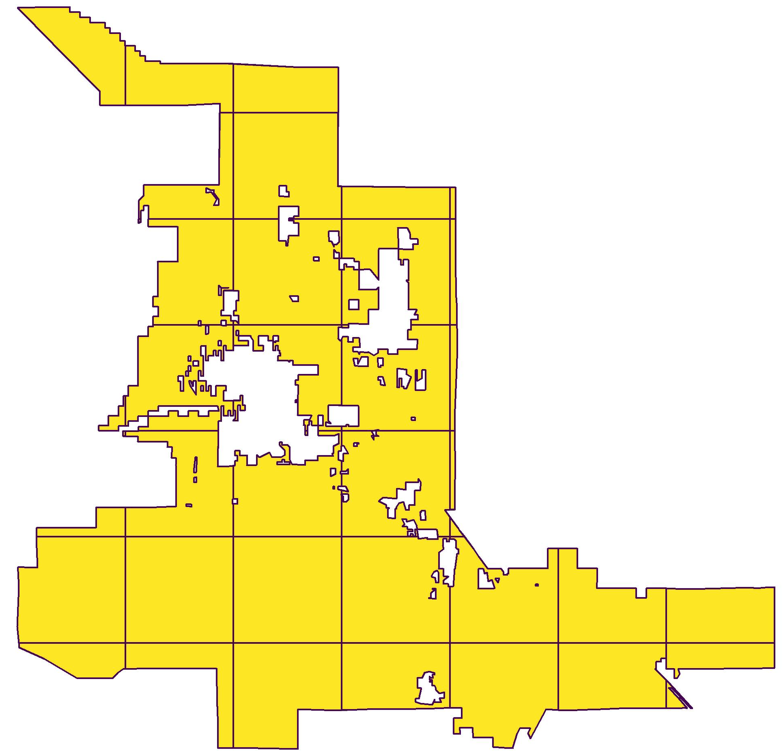}}
    \subfloat[Toronto]{
		\includegraphics[width=0.33\textwidth]{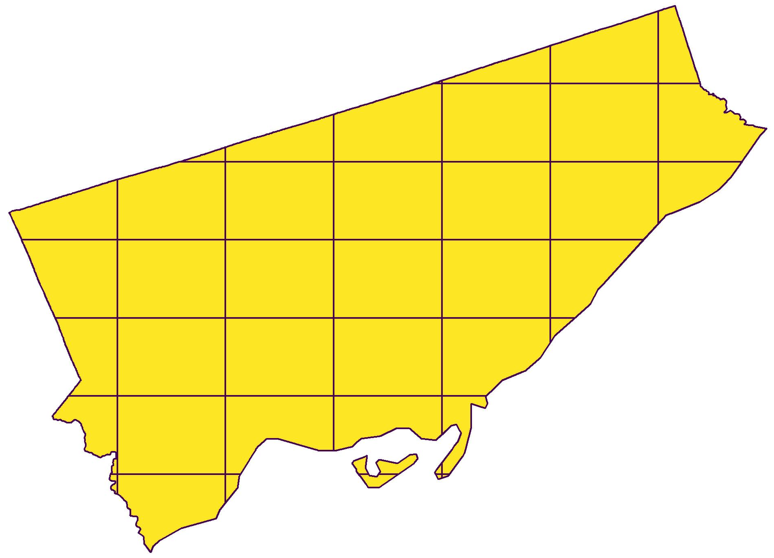}}
    \subfloat[Phoenix]{
    		\includegraphics[width=0.18\textwidth]{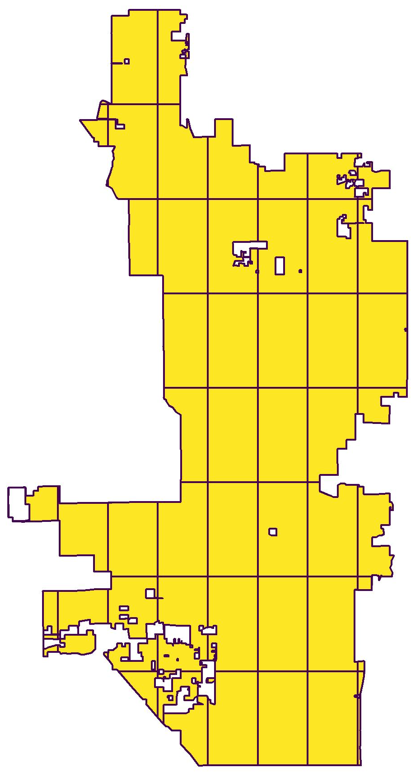}}\hfill
    \subfloat[Charlotte]{
    		\includegraphics[width=0.29\textwidth]{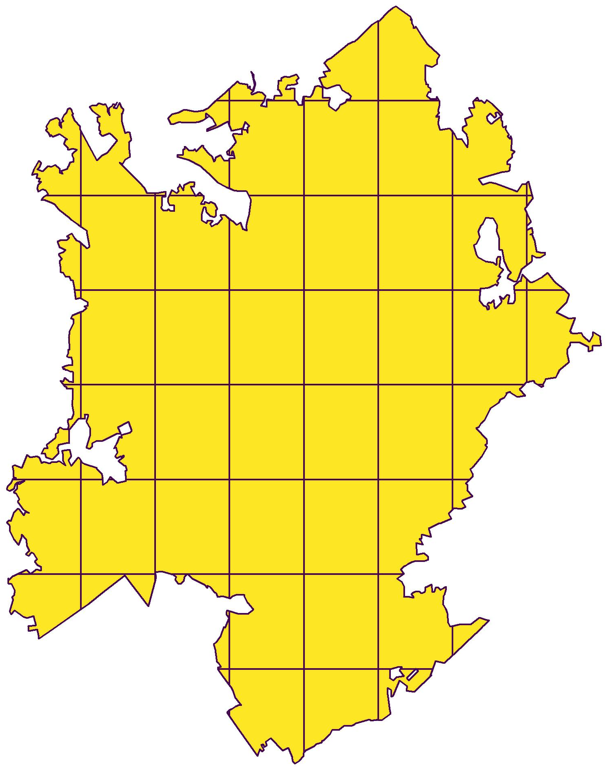}}
    \subfloat[Scottsdale]{
    		\includegraphics[width=0.15\textwidth]{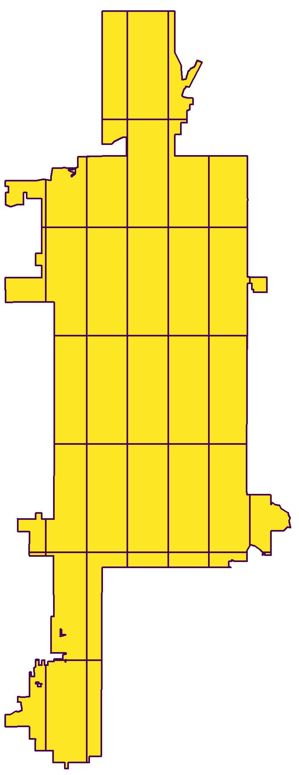}}
	\caption{The geographical map of divided districts in each
          city. We set $G=7$ for Las Vegas, Toronto and Scottsdale,
          and $G=8$ for Phoenix and Charlotte.
}
	\label{f:map}
\end{figure}

We focus on the commenting behaviour of users and try to investigate
the influential factors that can help predict a user's commenting
behavior. The analysis is conducted at the city
level. Specifically, we select the top five cities (i.e., Las Vegas,
Toronto, Phoenix, Charlotte, and Scottsdale) as examples, which have the
most business shops in the data set. Then in each city, we classify all
shops in the city into  districts. To this end, we first
sort the longitudes of all shops in an increasing order and then
evenly divide them into $G$ parts. The same operation is conducted for
the latitudes of all shops. This leads to  $N_2\le G^2$
districts in total. Figure \ref{f:map} illustrates the divided
districts in each city. The spatial adjacency matrix $\bA_2$ is then
constructed for the $N_2$ districts in a city. Specifically, we define
$a_{2ij} = 1$ if district $i$ and district $j$ are neighbors;
otherwise we have $a_{2ij} = 0$.
Subsequently we conduct a basic data cleaning procedure to keep active
users with total number of comments larger than 20.
We then construct  a friendship network $\bA_1$  among the active
users. Specifically we set $a_{1ij} = 1$ if user $i$ is
a friend of user $j$; otherwise $a_{1ij} = 0$.

The detailed description of the final data set in each city is
summarized in \textit{panel A} of Table \ref{t:t1}. To characterize
the network structure in each city, we compute the network density for
$\bA_1$ and $\bA_2$, i.e., $\sum_i\sum_ja_{1ij}/N_1^2$ and
$\sum_i\sum_ja_{2ij}/N_2^2$.
As shown by Table \ref{t:t1}, the user network $\bA_1$ is quite sparse
in all cities. In each city, we define the response variable $Y_{ijt}$ as the
average score that user $i$ commenting on shops in district $j$ during year $t$. Following the common practice in recommendation systems \citep{2017Graph,mao2019matrix,2020Inductive}, we treat average scores of users as missing values. Specifically, let $R_{ijt}$ denote whether we observe user $i$
commenting on shops in district $j$ at time $t$. Then we can
only observe $Y_{ijt}$ when $R_{ijt}=1$. The observation rate in each
city is then calculated and reported in Table \ref{t:t1}. As one can
see, all cities have low observation rates, indicating the whole
matrix in each city is sparse.
\begin{table}[h]
	\caption{Description of data and variables in five cities.}
	\label{t:t1}
	\begin{center}
		\begin{threeparttable}[b]
			\begin{tabular}{cccccc}
            \hline
            \hline
				City	&Las Vegas	&Toronto	&Phoenix	&Charlotte	&Scottsdale	\\
            \hline
                \multicolumn{6}{c}{\textit{Panel A: Basic Statistics of Data}}\\
            \hline
                $N_1$	&248	&269	&168	&101	&112	\\
                $N_2$	&49	      &49	&64	    &64	    &49	\\
                Time	&2010--2018	&2010--2018	&2010--2018	&2010--2018	&2010--2018	\\
                Density of $\bA_1$	&0.0022	&0.0055	&0.0074	&0.0131	&0.0061	\\
                Density of $\bA_2$	&0.1531	&0.1531	&0.1181	&0.1181	&0.1531	\\
                Observation Rate	&0.2388	&0.2932	&0.1833	&0.2813	&0.1980	\\
            \hline
                \multicolumn{6}{c}{\textit{Panel B: Mean of Variables}}\\
            \hline
                Duration	&0.3952	&0.6171	&0.6131	&0.4059	&0.6607	\\
                VIP\%	&0.4677	&0.1599	&0.3214	&0.1881	&0.3393	\\
                Useful	&1.8287	&0.9139	&2.0530	&0.8669	&1.9100	\\
                Funny	&1.2570	&0.3592	&1.3300	&0.3755	&1.1880	\\
                Cool	&1.4413	&0.5176	&1.3960	&0.4638	&1.3110	\\
                Average Score	&3.8477	&3.6765	&3.9724	&3.8973	&3.9339	\\
            \hline
            \hline
			\end{tabular}
		\end{threeparttable}
	\end{center}
\end{table}

\begin{figure}[h]
	\centering
	\subfloat[Average score per user vs. two duration levels]{
		\includegraphics[width=0.95\textwidth]{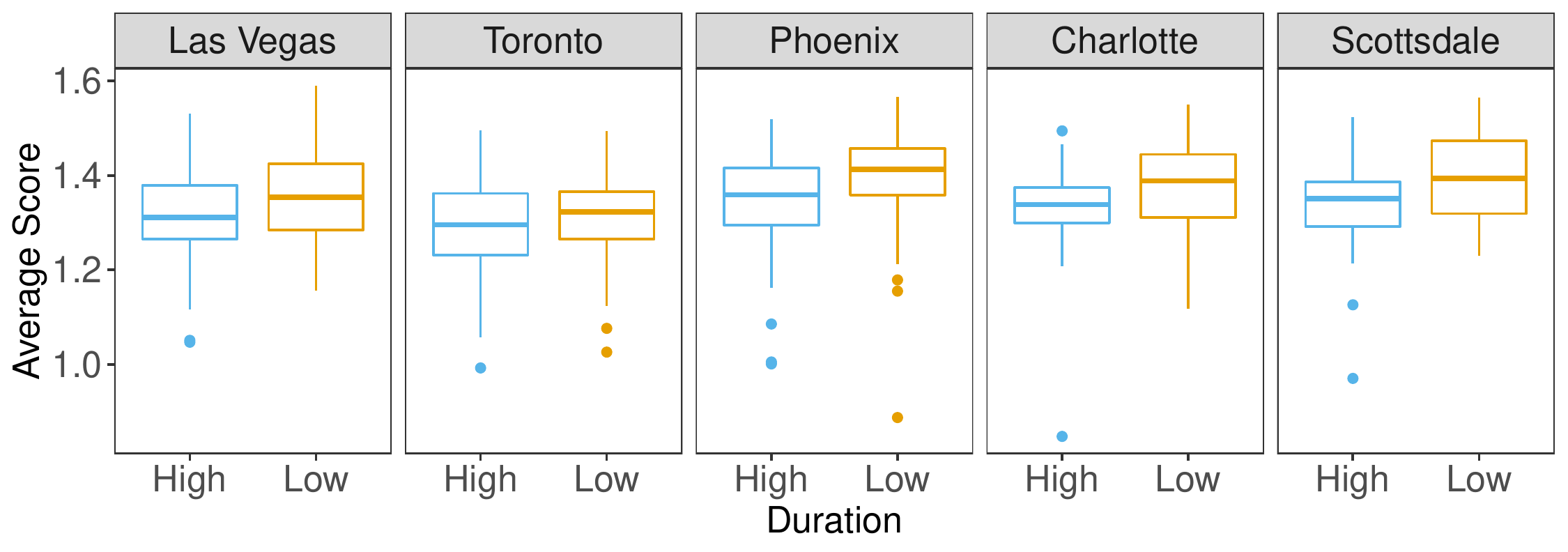}}\hfill
    \subfloat[Average score per user vs. VIP or non-VIP]{
		\includegraphics[width=0.95\textwidth]{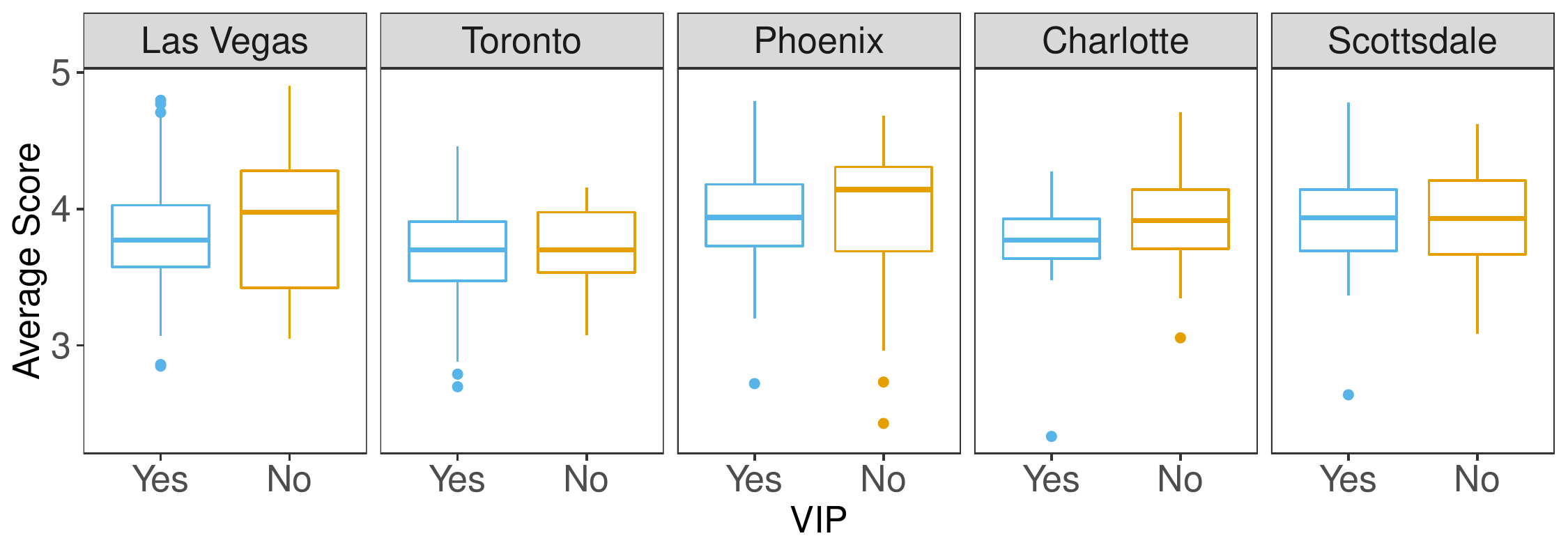}}\hfill
	\caption{The boxplots of average score per user in
          different duration or VIP groups.}
	\label{f:desp}
\end{figure}

We consider five user-specific covariates in the analysis, which
are computed based on the user information up to the year
2010. Specifically, the first one is the number of years from the
user's first registration to the year 2010, which we denoted by
``duration''. The second one is whether the user is VIP or not, which
is encoded by 1 and 0, respectively. The last three covariates are the
cumulated number of ``useful'', ``cool'', and ``funny'' comments given by
the users. For the observed response and all five covariates, we compute their
mean values in each city, which are summarized in \textit{panel B} of
Table \ref{t:t1}.
We also explore the relationship between the response and each
covariate. To this end, we first calculate
the average score of the posted comments by each user in all districts
during the whole time span, and then investigate the distributions of the
average score per user under different covariates. For illustration,
Figure \ref{f:desp} presents the boxplots of average score per user
under different duration groups (split by its median value) and VIP
groups in five cities. By Figure \ref{f:desp}, we find users with
shorter registration time tend to give higher scores in all five
cities. In addition, for cities Las Vegas, Phoenix, and Charlotte, VIP
users tend to give lower scores than non-VIP users.

We then investigate the rating behaviour of users in each city
using the MNAR method. For comparison purpose, the SVT-type methods
(i.e., SVT-AVG and SVT-SUM) \citep{mao2019matrix} are considered as
competitors. The SVT-SEP method is not considered, given its poor
performance in the simulation studies.
For the MNAR method, we conduct the bias reduction operation for two
rounds considering the time span is relatively short. The MNAR-ORG
method is also implemented for comparison purpose.
We consider two missing mechanisms MAR and UNI respectively. Under the missing
mechanism of MAR, the observation probability $p_i$ is estimated via a
logistic regression model with the five user-specific covariates
described above. Under the missing mechanism of UNI, the empirical
observation probabilities shown in Table \ref{t:t1} are used
directly.
To select the tuning parameters, we split the whole data set in each
city into three parts: (1) the training data set during $t=1$ to $T-2$,
(2) the validation data set at time $T-1$, and (3) the test data set at
time $T$. We first estimate each model on the training data set with
different values of tuning parameters. Then we evaluate the prediction
performance at the validation data set to choose the best tuning
parameters. To evaluate the prediction performance, we calculate
RMSE using the observed responses and its corresponding predicted
values in the validation data set. After the tuning parameters are
chosen, the training and validation data sets are combined
together to train a final model, and the test data in the last year
are used for method comparison.

The detailed results of RMSE under two missing mechanisms are
summarized in Table \ref{t:results}. Under each missing mechanism, the
proposed MNAR method outperforms the SVT-type methods in all five
cities by achieving lower RMSE values.
In addition, for all cities, the MNAR-ADJ method has obtained better
prediction performance than the MNAR-ORG method. It again demonstrates
the effectiveness of the bias reduction operation. Comparing different
missing mechanisms, we find the RMSE results under the missing
mechanism of MAR are all smaller than those under the missing
mechanism of UNI.
This result indicates that the MAR missing mechanism, which takes covariate information into
account, is necessary for the analytical task of this data set.

\begin{table}[h]
	\caption{The out-sample RMSE values obtained by different methods in five cities}
	\label{t:results}
	\begin{center}
		\begin{threeparttable}[b]
			\begin{tabular}{ccccccc}
            \hline
            \hline
				&City	&Las Vegas	&Toronto	&Phoenix	&Charlotte	&Scottsdale	\\
            \hline
            \multirow{4}{*}{MAR}
                &	MNAR-ORG	&0.5703	&0.4319	&0.4445	&0.4821	&0.5300	\\
                &	MNAR-ADJ	&0.5011	&0.3644	&0.3846	&0.4149	&0.4537	\\
                &	SVT-AVG	&0.6191	&0.4392	&0.4815	&0.5239	&0.5725	\\
                &	SVT-SUM	&0.6239	&0.4646	&0.4907	&0.5386	&0.5829	\\
            \hline
            \multirow{4}{*}{UNI}
                &	MNAR-ORG	&0.9258	&0.6617	&1.1018	&0.7363	&1.0626	\\
                &	MNAR-ADJ	&0.7362	&0.6125	&0.9936	&0.6153	&0.9442	\\
                &	SVT-AVG	&1.1161	&0.7077	&1.2980	&0.8326	&1.3058	\\
                &	SVT-SUM	&1.1370	&0.7779	&1.3291	&0.8853	&1.3299	\\
            \hline
            \hline
			\end{tabular}
		\end{threeparttable}
	\end{center}
\end{table}

\begin{figure}[h]
	\centering
	\subfloat[$\widehat{\lambda}_i$ with $1\leq i \leq N_1$]{
		\includegraphics[width=0.95\textwidth]{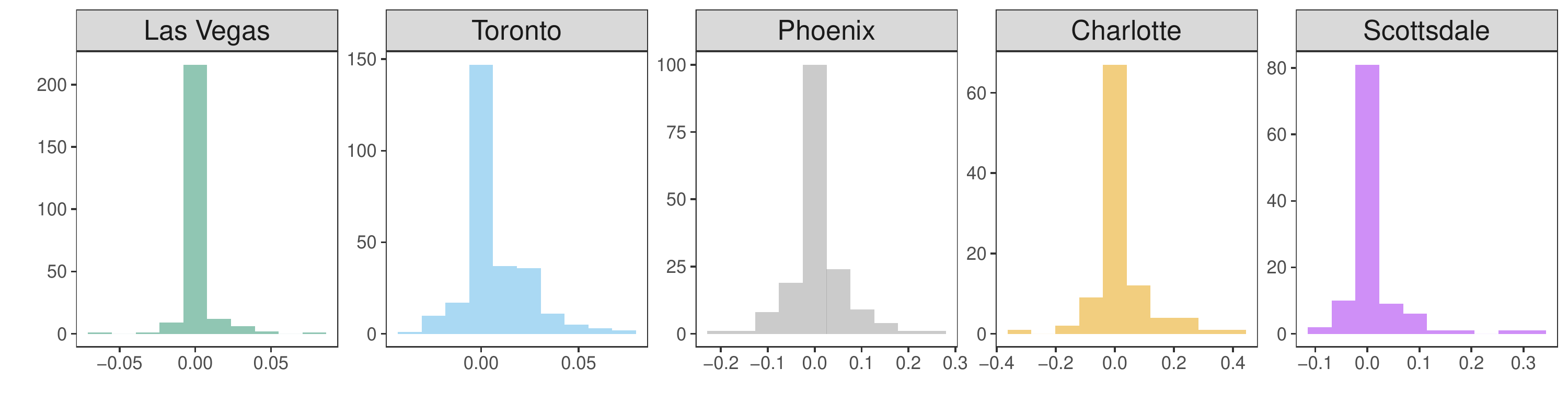}}\hfill
    \subfloat[$\widehat{\gamma}_j$ with $1\leq j \leq N_2$]{
		\includegraphics[width=0.95\textwidth]{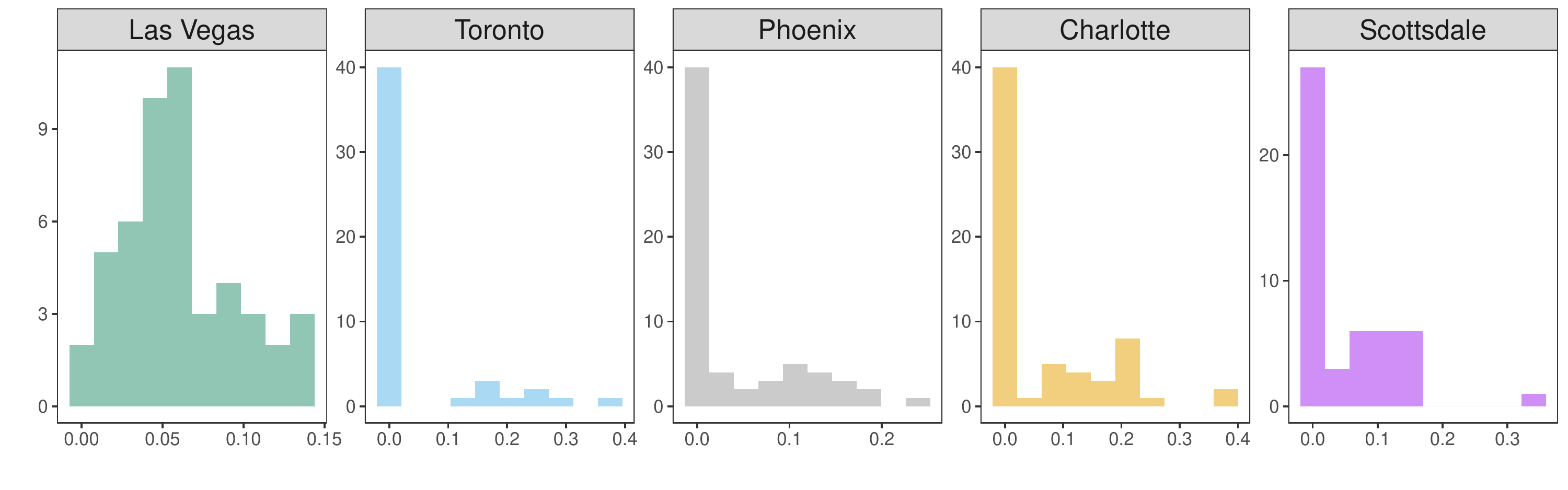}}\hfill
	\caption{The histograms of estimated user-specific effects $\widehat{\lambda}_i$ and district-specific effects $\widehat{\gamma}_j$ in five cities.}
	\label{f:est}
\end{figure}

Next, we focus on the estimated network effects. We consider for
example the estimates of user-specific effects (i.e.,
$\widehat{\lambda}_i$) and district-specific effects (i.e.,
$\widehat{\gamma}_j$) by MNAR-ADJ under the missing mechanism of
MAR. To illustrate the network effects, Figure \ref{f:est} shows the
estimated user-specific effects and district-specific
effects in five cities.
As shown, the estimated values of $\widehat{\lambda}_i$ and
$\widehat{\gamma}_j$ vary in different cities. In general, most of the
estimated user-specific effects $\widehat{\lambda}_i$ vary between
-0.4 to 0.4. This finding suggests that, the influences from friends
on users' commenting behaviors can be either negative or positive.
In contrast, the estimated district-specific effects
$\widehat{\gamma}_j$ are all non-negative.
This indicates an overall positive influence from neighboring business.
Moreover, the skewed distribution of $\widehat{\gamma}_j$s reflects
the asymmetric influences of different districts.

Finally, we test the significance of the estimated network effects.
As shown by Figure \ref{f:est}, many users or districts have very
small network effects. By setting a threshold $\delta=0.05$, we regard
users to be active if their corresponding user-specific effects are
larger than the threshold. Specifically, define
$\Delta_{u}=\{i:\widehat{\lambda}_i>\delta\}$ to be the set of active
users. Similarly, define $\Delta_d=\{j:\widehat{\gamma}_j>\delta\}$ to
be the set of active districts. We then compute the averaged
user-specific effect for all active users as $\bar{\lambda}=\sum_{i
  \in \Delta_{u}}\widehat{\lambda}_i/|\Delta_{u}|$, where
$|\Delta_{u}|$ denotes the total number of active users. Similarly,
the averaged district-specific effect for all active districts can be
computed as $\bar{\gamma}=\sum_{j \in
  \Delta_{d}}\widehat{\gamma}_j/|\Delta_{d}|$. Then we test the
significance of $\bar{\lambda}$ and $\bar{\gamma}$ according to
Theorem \ref{thm:bias_corrections}. Specifically, to calculate the asymptotical variance of
$\bar{\lambda}$, the vector $\bm{\eta}$ used in Theorem \ref{thm:bias_corrections} is specified
as follows. Let
$\bm{\eta}_1=\{\eta_{11},...,\eta_{1N_1}\}^{\top}\in\mR^{N_1}$. Then
define $\eta_{1i}=1/|\Delta_{u}|$ if $i \in \Delta_{u}$, otherwise
$\eta_{1i}=0$. Then
$\bm{\eta}=(\bm{\eta}_1^{\top},\bm{0}_{N_2})^{\top}$. Similarly, let
$\bm{\eta}_2=\{\eta_{21},...,\eta_{2N_2}\}^{\top}\in\mR^{N_2}$, where
$\eta_{2j}=1/|\Delta_{d}|$ if $j \in \Delta_{d}$, otherwise
$\eta_{2j}=0$. Then, to calculate the asymptotical variance of
$\bar{\gamma}$, the vector $\bm{\eta}$ used in Theorem
\ref{thm:bias_corrections} is specified as
$\bm{\eta}=(\bm{0}_{N_1},\bm{\eta}_2^{\top})^{\top}$.

\begin{figure}[h]
	\centering
	\subfloat[Averaged User-Specific Effect]{
		\includegraphics[width=0.48\textwidth]{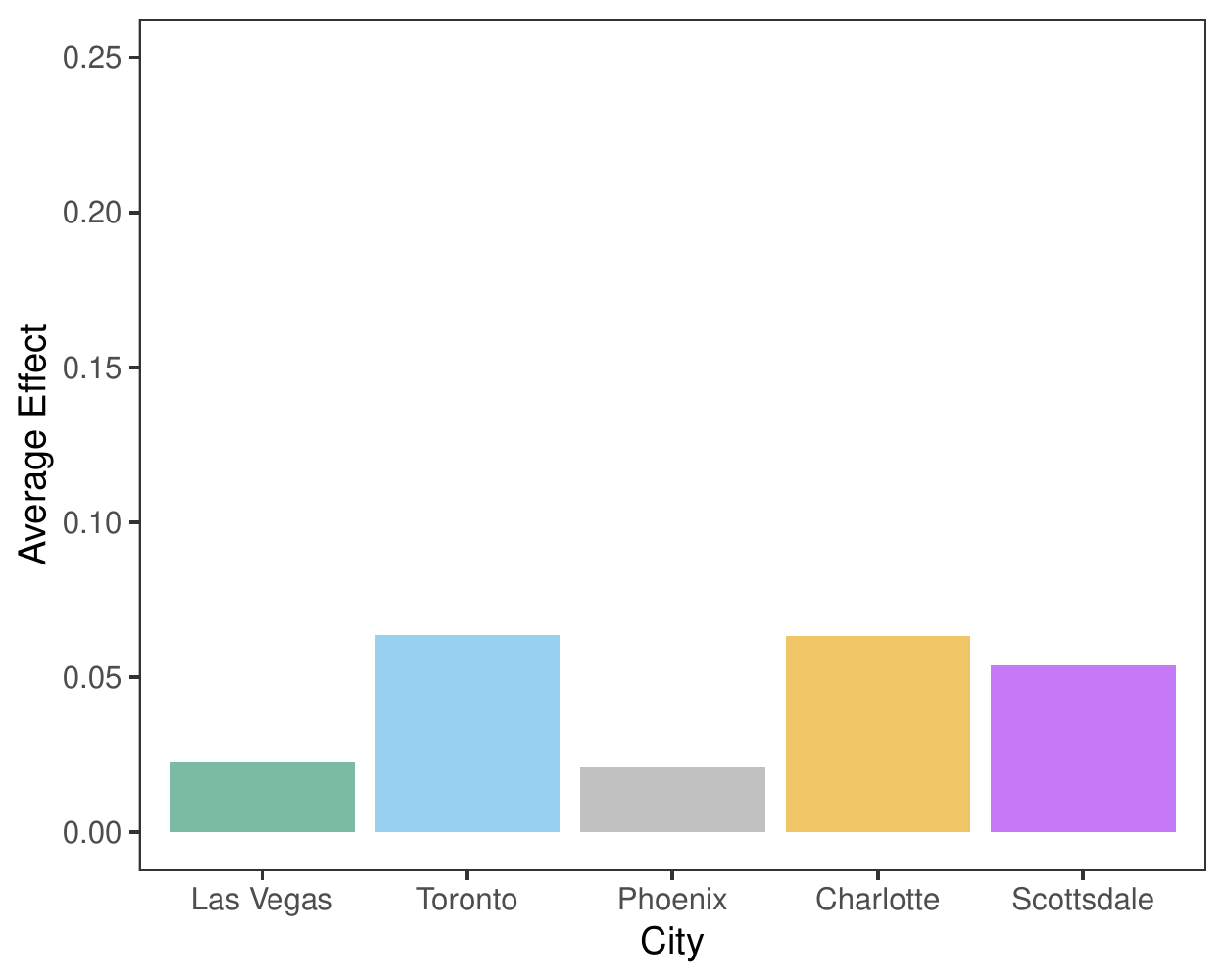}}\hfill
    \subfloat[Averaged District-Specific Effect]{
		\includegraphics[width=0.48\textwidth]{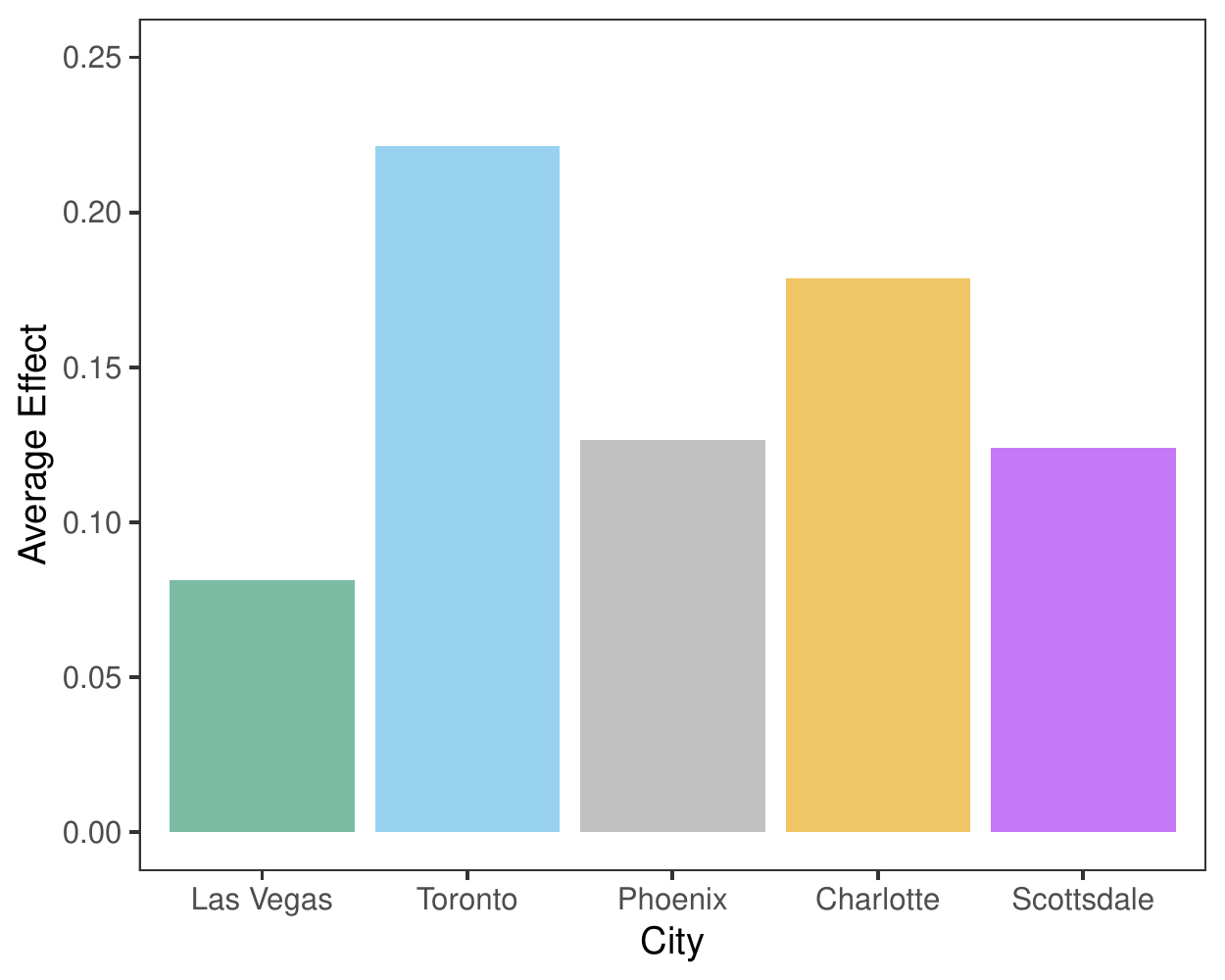}}\hfill
	\caption{The averaged user-specific effects ($\bar{\lambda}$) and district-specific effects ($\bar{\gamma}$) in five cities.}
	\label{f:sig}
\end{figure}

Figure \ref{f:sig} presents the averaged user-specific effect
$\bar{\lambda}$ and averaged district-specific effect $\bar{\gamma}$
in each city.
Under the significance level 5\%, we find all $\bar{\lambda}$s and
$\bar{\gamma}$s are significant.
Specifically, the averaged user-specific effects in the five cities
are relatively small, with all values smaller than 0.1.
Compared with the averaged user-specific effects, the averaged
district-specific effects are larger. This finding suggests that, the
behaviours of users in the Yelp platform are more likely to be
influenced by the visited location than their friends. In addition,
among the five cities, Toronto and Charlotte have more obvious user-specific effects and district-specific
effects than the other three cities. These results again demonstrate the
heterogeneous characteristics of different cities.

\section{Concluding Remarks}
\label{sec:conclude}

In this work we propose a matrix network autoregression model, which
accommodates incomplete matrix observations.
We now discuss some potential extensions to our work, which may be
interesting research topics for future studies.
First, to better characterize the dynamics of the matrix-valued time
series data,
matrix factor structure \citep{wang2019factor,chen2021statistical} can
be further considered and studied.
Second, non-linear and non-parametric modelling frameworks can be incorporated
to explore more flexible dynamics of the matrix-valued time series data.
Third, more refined treatment \citep{tsiatis2006}
can be incorporated to handle the missing response, which may improve
the performance especially when $c_p$ is small. On this hand,  it may be of
interest to consider the situation when the missingness further
depends on the response itself, which is much more challenging but can
be important in certain applications.

\newpage

\bibliographystyle{asa}
\bibliography{xuening}

\newpage
\begin{landscape}
\begin{table}[!h]
\caption{Simulation results for estimation performance under the missing mechanism of MAR. The empirical root mean square errors (RMSE) for $(\bLambda,\bGamma,\bbeta,\bB,\bA)$, as well as the test errors under different methods are reported, respectively. For illustration purpose, the RMSEs for $\bbeta$  are multiplied by 100.}\label{t:case1}
\begin{center}
\small
\renewcommand\arraystretch{1}
\begin{threeparttable}[b]
\begin{tabular}{cccccccccccccccccccc}
\hline
\hline
\multirow{2}{*}{$N_1$} &\multirow{2}{*}{$N_2$} &\multirow{2}{*}{$\bm{\theta}$} &\multicolumn{3}{c}{SVT} &\multicolumn{2}{c}{MNAR} & &\multicolumn{3}{c}{SVT} &\multicolumn{2}{c}{MNAR} & &\multicolumn{3}{c}{SVT} &\multicolumn{2}{c}{MNAR}\\
\cline{4-8}
\cline{10-14}
\cline{16-20}
&&	&SEP	&AVG	&SUM &ORG  &ADJ	&   &SEP	&AVG	&SUM &ORG  &ADJ	& &SEP	&AVG	&SUM &ORG  &ADJ	\\
\cline{4-20}
&& &\multicolumn{5}{c}{$T=30$} & &\multicolumn{5}{c}{$T=60$} & &\multicolumn{5}{c}{$T=100$}\\
\multirow{6}{*}{100} &\multirow{6}{*}{100}
&$\bLambda$	&$-$	&$-$	&$-$	&0.383	&0.337	&	&$-$	&$-$	&$-$	&0.331	&0.282	&	&$-$	&$-$	&$-$	&0.280	&0.230	\\
&&$\bGamma$	&$-$	&$-$	&$-$	&0.385	&0.340	&	&$-$	&$-$	&$-$	&0.331	&0.281	&	&$-$	&$-$	&$-$	&0.284	&0.234	\\
&&$\bbeta$	&4.577	&1.309	&1.282	&1.153	&0.996	&	&4.581	&1.166	&1.153	&0.906	&0.765	&	&4.581	&1.104	&1.096	&0.733	&0.577	\\
&&$\bB$	&1.619	&0.260	&0.263	&0.255	&0.254	&	&1.619	&0.202	&0.203	&0.184	&0.180	&	&1.619	&0.175	&0.175	&0.146	&0.141	\\
&&$\bA$	&2.650	&0.310	&0.313	&0.302	&0.295	&	&2.651	&0.264	&0.265	&0.244	&0.235	&	&2.650	&0.245	&0.245	&0.213	&0.204	\\
&&Error	&1.491	&0.158	&0.162	&0.151	&0.143	&	&1.492	&0.116	&0.117	&0.098	&0.091	&	&1.492	&0.099	&0.099	&0.075	&0.069	\\
\hline
\multirow{6}{*}{200} &\multirow{6}{*}{200}
&$\bLambda$	&$-$	&$-$	&$-$	&0.335	&0.295	&	&$-$	&$-$	&$-$	&0.268	&0.224	&	&$-$	&$-$	&$-$	&0.217	&0.183	\\
&&$\bGamma$	&$-$	&$-$	&$-$	&0.348	&0.303	&	&$-$	&$-$	&$-$	&0.286	&0.240	&	&$-$	&$-$	&$-$	&0.232	&0.194	\\
&&$\bbeta$	&1.741	&0.371	&0.366	&0.347	&0.333	&	&1.742	&0.290	&0.288	&0.260	&0.252	&	&1.742	&0.249	&0.248	&0.215	&0.209	\\
&&$\bB$	&1.211	&0.221	&0.223	&0.199	&0.194	&	&1.211	&0.187	&0.188	&0.145	&0.138	&	&1.211	&0.172	&0.173	&0.114	&0.108	\\
&&$\bA$	&2.713	&0.239	&0.241	&0.232	&0.231	&	&2.713	&0.209	&0.209	&0.195	&0.193	&	&2.713	&0.196	&0.196	&0.178	&0.176	\\
&&Error	&1.246	&0.089	&0.090	&0.084	&0.083	&	&1.246	&0.068	&0.068	&0.059	&0.058	&	&1.246	&0.060	&0.060	&0.049	&0.048	\\
\hline
\multirow{6}{*}{400} &\multirow{6}{*}{400}
&$\bLambda$	&$-$	&$-$	&$-$	&0.293	&0.247	&	&$-$	&$-$	&$-$	&0.222	&0.189	&	&$-$	&$-$	&$-$	&0.174	&0.152	\\
&&$\bGamma$	&$-$	&$-$	&$-$	&0.283	&0.239	&	&$-$	&$-$	&$-$	&0.212	&0.182	&	&$-$	&$-$	&$-$	&0.165	&0.145	\\
&&$\bbeta$	&0.721	&0.281	&0.275	&0.212	&0.185	&	&0.721	&0.267	&0.264	&0.173	&0.153	&	&0.721	&0.260	&0.258	&0.152	&0.140	\\
&&$\bB$	&0.791	&0.176	&0.178	&0.145	&0.140	&	&0.791	&0.156	&0.156	&0.104	&0.101	&	&0.791	&0.147	&0.147	&0.081	&0.079	\\
&&$\bA$	&2.733	&0.228	&0.230	&0.204	&0.197	&	&2.733	&0.214	&0.214	&0.176	&0.172	&	&2.733	&0.207	&0.207	&0.163	&0.161	\\
&&Error	&1.193	&0.081	&0.082	&0.065	&0.060	&	&1.194	&0.071	&0.071	&0.049	&0.046	&	&1.193	&0.067	&0.067	&0.042	&0.040	\\
\hline	
\multirow{6}{*}{600} &\multirow{6}{*}{600}															
&$\bLambda$	&$-$	&$-$	&$-$	&0.250	&0.211	&	&$-$	&$-$	&$-$	&0.180	&0.158	&	&$-$	&$-$	&$-$	&0.138	&0.125	\\
&&$\bGamma$	&$-$	&$-$	&$-$	&0.248	&0.209	&	&$-$	&$-$	&$-$	&0.180	&0.158	&	&$-$	&$-$	&$-$	&0.138	&0.125	\\
&&$\bbeta$	&0.396	&0.159	&0.158	&0.147	&0.144	&	&0.396	&0.151	&0.150	&0.137	&0.136	&	&0.396	&0.148	&0.147	&0.133	&0.132	\\
&&$\bB$	&0.611	&0.160	&0.161	&0.119	&0.115	&	&0.611	&0.145	&0.145	&0.084	&0.082	&	&0.611	&0.138	&0.138	&0.065	&0.064	\\
&&$\bA$	&2.735	&0.194	&0.195	&0.179	&0.177	&	&2.735	&0.182	&0.183	&0.161	&0.160	&	&2.735	&0.177	&0.178	&0.153	&0.152	\\
&&Error	&1.108	&0.059	&0.060	&0.050	&0.049	&	&1.108	&0.052	&0.052	&0.041	&0.040	&	&1.108	&0.049	&0.049	&0.037	&0.036	\\
\hline
\hline
\end{tabular}
\end{threeparttable}
\end{center}
\end{table}

\begin{table}[!h]
\caption{Simulation results for estimation performance under the missing mechanism of UNI. The empirical root mean square errors (RMSE) for $(\bLambda,\bGamma,\bbeta,\bB,\bA)$, as well as the test errors under different methods are reported, respectively. For illustration purpose, the RMSEs for $\bbeta$  are multiplied by 100.}\label{t:case2}
\begin{center}
\small
\renewcommand\arraystretch{1}
\begin{threeparttable}[b]
\begin{tabular}{cccccccccccccccccccc}
\hline
\hline
\multirow{2}{*}{$N_1$} &\multirow{2}{*}{$N_2$} &\multirow{2}{*}{$\bm{\theta}$} &\multicolumn{3}{c}{SVT} &\multicolumn{2}{c}{MNAR} & &\multicolumn{3}{c}{SVT} &\multicolumn{2}{c}{MNAR} & &\multicolumn{3}{c}{SVT} &\multicolumn{2}{c}{MNAR}\\
\cline{4-8}
\cline{10-14}
\cline{16-20}
&&	&SEP	&AVG	&SUM &ORG  &ADJ	&   &SEP	&AVG	&SUM &ORG  &ADJ	& &SEP	&AVG	&SUM &ORG  &ADJ	\\
\hline
&& &\multicolumn{5}{c}{$T=30$} & &\multicolumn{5}{c}{$T=60$} & &\multicolumn{5}{c}{$T=100$}\\
\multirow{6}{*}{100} &\multirow{6}{*}{100}
&$\bLambda$	&$-$	&$-$	&$-$	&0.383	&0.337	&	&$-$	&$-$	&$-$	&0.331	&0.282	&	&$-$	&$-$	&$-$	&0.280	&0.230	\\
&&$\bGamma$	&$-$	&$-$	&$-$	&0.385	&0.340	&	&$-$	&$-$	&$-$	&0.331	&0.281	&	&$-$	&$-$	&$-$	&0.284	&0.234	\\
&&$\bbeta$	&4.573	&1.308	&1.280	&1.152	&0.996	&	&4.577	&1.165	&1.152	&0.905	&0.765	&	&4.576	&1.103	&1.095	&0.733	&0.577	\\
&&$\bB$	&1.618	&0.260	&0.263	&0.255	&0.254	&	&1.618	&0.202	&0.203	&0.184	&0.180	&	&1.618	&0.175	&0.175	&0.146	&0.141	\\
&&$\bA$	&2.647	&0.310	&0.313	&0.302	&0.295	&	&2.648	&0.264	&0.265	&0.244	&0.235	&	&2.648	&0.245	&0.245	&0.213	&0.204	\\
&&Error	&1.490	&0.158	&0.162	&0.151	&0.143	&	&1.491	&0.116	&0.116	&0.098	&0.091	&	&1.491	&0.099	&0.099	&0.075	&0.069	\\
\hline
\multirow{6}{*}{200} &\multirow{6}{*}{200}
&$\bLambda$	&$-$	&$-$	&$-$	&0.335	&0.295	&	&$-$	&$-$	&$-$	&0.268	&0.224	&	&$-$	&$-$	&$-$	&0.217	&0.183	\\
&&$\bGamma$	&$-$	&$-$	&$-$	&0.348	&0.303	&	&$-$	&$-$	&$-$	&0.286	&0.240	&	&$-$	&$-$	&$-$	&0.232	&0.194	\\
&&$\bbeta$	&1.741	&0.371	&0.366	&0.347	&0.333	&	&1.742	&0.290	&0.288	&0.260	&0.252	&	&1.742	&0.249	&0.248	&0.215	&0.209	\\
&&$\bB$	&1.211	&0.221	&0.223	&0.199	&0.194	&	&1.211	&0.187	&0.188	&0.145	&0.138	&	&1.211	&0.172	&0.173	&0.114	&0.108	\\
&&$\bA$	&2.712	&0.239	&0.241	&0.232	&0.231	&	&2.712	&0.209	&0.209	&0.195	&0.193	&	&2.712	&0.196	&0.196	&0.178	&0.176	\\
&&Error	&1.246	&0.088	&0.090	&0.084	&0.083	&	&1.246	&0.068	&0.068	&0.059	&0.058	&	&1.246	&0.060	&0.060	&0.049	&0.048	\\
\hline
\multirow{6}{*}{400} &\multirow{6}{*}{400}
&$\bLambda$	&$-$	&$-$	&$-$	&0.293	&0.247	&	&$-$	&$-$	&$-$	&0.222	&0.189	&	&$-$	&$-$	&$-$	&0.174	&0.152	\\
&&$\bGamma$	&$-$	&$-$	&$-$	&0.283	&0.239	&	&$-$	&$-$	&$-$	&0.212	&0.182	&	&$-$	&$-$	&$-$	&0.165	&0.145	\\
&&$\bbeta$	&0.721	&0.281	&0.275	&0.212	&0.185	&	&0.721	&0.267	&0.264	&0.173	&0.153	&	&0.721	&0.260	&0.258	&0.152	&0.140	\\
&&$\bB$	&0.791	&0.176	&0.178	&0.145	&0.140	&	&0.791	&0.156	&0.156	&0.104	&0.101	&	&0.791	&0.147	&0.147	&0.081	&0.079	\\
&&$\bA$	&2.733	&0.228	&0.230	&0.204	&0.197	&	&2.733	&0.214	&0.214	&0.176	&0.172	&	&2.733	&0.207	&0.207	&0.163	&0.161	\\
&&Error	&1.193	&0.081	&0.082	&0.065	&0.060	&	&1.194	&0.071	&0.071	&0.049	&0.046	&	&1.193	&0.067	&0.067	&0.042	&0.040	\\
\hline	
\multirow{6}{*}{600} &\multirow{6}{*}{600}															
&$\bLambda$	&$-$	&$-$	&$-$	&0.250	&0.211	&	&$-$	&$-$	&$-$	&0.180	&0.158	&	&$-$	&$-$	&$-$	&0.138	&0.125	\\
&&$\bGamma$	&$-$	&$-$	&$-$	&0.248	&0.209	&	&$-$	&$-$	&$-$	&0.180	&0.158	&	&$-$	&$-$	&$-$	&0.138	&0.125	\\
&&$\bbeta$	&0.396	&0.159	&0.158	&0.147	&0.144	&	&0.396	&0.151	&0.150	&0.137	&0.136	&	&0.396	&0.148	&0.147	&0.133	&0.132	\\
&&$\bB$	&0.611	&0.160  &0.161	&0.119	&0.115	&	&0.611	&0.145	&0.145	&0.084	&0.082	&	&0.611	&0.138	&0.138	&0.065	&0.064	\\
&&$\bA$	&2.735	&0.194	&0.195	&0.179	&0.177	&	&2.735	&0.182	&0.183	&0.161	&0.160	&	&2.735	&0.177	&0.178	&0.153	&0.152	\\
&&Error	&1.108	&0.059	&0.060	&0.050	&0.049	&	&1.108	&0.052	&0.052	&0.041	&0.040	&	&1.108	&0.049	&0.049	&0.037	&0.036	\\
\hline
\hline
\end{tabular}
\end{threeparttable}
\end{center}
\end{table}
\end{landscape}

\end{document}